\documentclass[12pt]{article}

\usepackage[UKenglish]{babel}

\usepackage{latexsym}
\usepackage{amssymb}
\usepackage{epsfig}

\begin{document}

\title{The band-centre anomaly in the 1D Anderson model with
correlated disorder}

\author{L. Tessieri${}^{1}$~\footnote{Corresponding author:
luca.tessieri@gmail.com}, I. F. Herrera-Gonz\'{a}lez${}^{2}$,
F. M. Izrailev${}^{2}$,\\
{\it ${}^{1}$ Instituto de F\'{\i}sica y Matem\'{a}ticas} \\
{\it Universidad Michoacana de San Nicol\'{a}s de Hidalgo} \\
{\it 58060, Morelia, Mexico} \\
{\it ${}^{2}$ Instituto de F\'{\i}sica, Universidad Aut\'{o}noma
de Puebla,} \\
{\it Puebla, 72570, Mexico}}

\date{3rd June 2015}

\maketitle

\begin{abstract}
We study the band-centre anomaly in the one-dimensional Anderson model
with weak {\em correlated} disorder. Our analysis is based on the
Hamiltonian map approach; the correspondence between the discrete model
and its continuous counterpart is discussed in detail. We obtain
analytical expressions of the localisation length and of the invariant
measure of the phase variable, valid for energies in a neighbourhood of
the band centre.
By applying these general results to specific forms of correlated disorder,
we show how correlations can enhance or suppress the anomaly at the band
centre.
\end{abstract}

Pacs: 71.23.An, 72.15.Rn, 05.40.-a

\section{Introduction}

The one-dimensional (1D) Anderson model, first introduced in 1958~\cite{And58},
remains the focus of active research. The analysis of this model showed that
all eigenstates are exponentially localised for infinitely large samples
(see, e.g., the reviews~\cite{Lee85,Ben97}).
For the case of weak and uncorrelated disorder, Thouless derived the
expression for the localisation length that now bears his name~\cite{Tho79}.
Numerical calculations, however, revealed that Thouless' expression
fails at the band centre, where the actual localisation length
was numerically found to be different from the expected value~\cite{Czy81}.
As was understood by Kappus and Wegner~\cite{Kap81}, this discrepancy
is due to the failure of the Born approximation; with the use of a quite
sophisticated method, Kappus and Wegner derived an approximate expression
which explained the numerical data reported in Ref.~\cite{Czy81}. 
Later on, using a different approach, Derrida and Gardner showed~\cite{Der84}
that this anomaly is a resonance effect that emerges at the band centre,
$E=0$. They also found another anomaly for $E= \pm 1$ and suggested that
similar anomalies should appear for other resonant energies defined by
$E = 2 \cos ( \pi p/q)$, where $p$ and $q$ are integer number.
The results of Ref.~\cite{Der84} were subsequently extended to a whole
neighbourhood of the band centre~\cite{Kus93,Gol94} (see also the discussion
in Ref.~\cite{Izr12}).

The interest in the anomalies of the Anderson model was raised significantly
after these anomalies were linked to the so-called single-parameter scaling
theory (SPS). The SPS theory was proposed in Ref.~\cite{Abr79, And80} and
was based on the random phase approximation (RPA) for the phases of the
scattered waves. In the 1D case, the SPS theory can be reduced to the
statement that all moments of the transmission coefficient through a random
barrier can be expressed in terms of the first two moments.
As a result, the whole distribution of the transmission coefficient is
expected to depend on the ratio between the localisation length and the
size of the random chain.
However, it was numerically shown long ago~\cite{Sto83} that the RPA fails
at the band centre of the Anderson model. Thus, the band centre anomaly
(as well as the other resonances) can serve as a touchstone for the study
of the SPS hypothesis. As was recently shown in Ref.~\cite{Sch03}, the SPS
hypothesis is violated also for the resonance at $E= \pm 1$ (see also the
discussion in Refs.~\cite{Dey00}). Since the phase distribution is not flat
whenever the energy lies in a whole neighbourhood of the resonant values,
one can expect that the SPS hypothesis should not be rigorously valid in
the whole energy band of the 1D Anderson model with weak disorder.
This ``anti-SPS'' hypothesis is still not proved; the study of the anomalies
of the Anderson model for $E=0$ and $E= \pm 1$ can provide a way to test the
validity of this conjecture.

The works on the band-centre anomaly mentioned so far considered the
Anderson model with {\em uncorrelated} disorder. In the literature, almost
no attention has been given to the case of correlated disorder, the main
exception being Ref.~\cite{Tit05} which, however, was focused on specific
cases of very short-ranged correlations.
Only very recently the effects of correlated disorder on the localisation
of the band-centre states have begun to be investigated, in the wake of the
discovery that the band-centre anomaly can be strongly reduced if the site
energies exhibit exponentially decaying, positive correlations~\cite{Her15}.

In this work we derive analytical results that show how the band-centre
anomaly is modified when disorder possesses {\em arbitrary} correlations.
The general expressions that we obtained allow us to treat the correlations
considered in~\cite{Her15} as a particular case and to show how, in general,
different kinds of disorder correlations can either enhance or suppress the
anomaly at the centre of the energy band.
To derive an analytical expression for the localisation length in a
neighbourhood of the band centre, we relied on the Hamiltonian map
formalism~\cite{Izr95, Izr12} and we replaced a map for the angle variable
with its continuum limit. In order to do so, we had to establish a rigorous
correspondence between random maps with intercorrelated coloured noises
and stochastic differential equations with the same features.
This led us to derive a discrete integration scheme for stochastic
differential equations with self- and cross-correlated noises.
This scheme is another important result of the present paper.

To analyse the band-centre anomaly we considered the behaviour of the
localisation length and that of the probability distribution for the angle
variable (which is equivalent to the scattering phase).
Our analytical results show that disorder correlations shape the localisation
length in a twofold way: on the one hand, the modulation of the phase
distribution causes a deviation of the actual Lyapunov exponent from the
value predicted by the expression first derived by Izrailev and
Krokhin~\cite{Izr99}, which generalises Thouless' formula to the case of
correlated disorder. This discrepancy is a resonance effect and represents
the ``anomaly'' in the presence of correlated disorder.
On the other hand, disorder correlations modify the localisation length via
the power-spectrum factor which is already present in the formula obtained
by Izrailev and Krokhin in Ref.~\cite{Izr99}.
By manipulating this factor, one can produce very strong localisation or
effective delocalisation of the band-centre states.

The modulation of the phase distribution close to the band centre is the
hallmark of the anomaly that occurs there.
Our analytical expressions show that positive, exponentially decaying
correlations of the disorder flatten the invariant distribution, thereby
reducing the anomaly, as observed in~\cite{Her15}.
The opposite effect occurs when correlations decrease exponentially in
magnitude but oscillate between positive and negative values.
Such correlations strongly enhance the band-centre anomaly.
To complete the picture, we consider a third type of correlations, which
describe a lattice formed by two statistically independent and
physically interpenetrating sublattices. We show that these correlations
do not alter the modulation of the invariant distribution with respect
to the case of uncorrelated disorder; however, they can increase or
decrease the energy interval over which the anomaly is significant.

This paper is organised as follows. In Sec.~\ref{hamapp} we define the model
under study and we introduce the Hamiltonian map approach.
In Sec.~\ref{rm_sde} we show how to replace the random map for the phase
variable with a corresponding stochastic differential equation. We proceed
to derive general analytical expressions for the phase distribution and the
localisation length in Sec.~\ref{general_formulae}.
After recovering the known results for the limit case of uncorrelated disorder
in Sec.~\ref{uncorrelated_disorder}, we consider the case of disorder with
positive and exponentially decreasing correlations in Sec.~\ref{exp_dec}.
The case of correlations oscillating between positive and negative values
and with exponentially decreasing magnitude is discussed in
Sec.~\ref{oscexpdec}.
A case of long-ranged correlations is analysed in Sec.~\ref{longrangedcor}.
We finally draw our conclusions in Sec.~\ref{conclu}.

\section{The Hamiltonian map approach}
\label{hamapp}

\subsection{Definition of the model}

We consider the 1D Anderson model with weak and correlated disorder.
The model is defined by the Schr\"{o}dinger equation
\begin{equation}
\psi_{n+1} + \psi_{n-1} + \varepsilon_{n} \psi_{n} = E \psi_{n} 
\label{andmod}
\end{equation}
with random site energies $\varepsilon_{n}$. We use energy units such
that $\hbar^{2}/2m =1$. We assume that
\begin{equation}
\begin{array}{ccc}
\langle \varepsilon_{n} \rangle = 0 &
\mbox{ and } &
\langle \varepsilon_{n}^{2} \rangle = \sigma^{2} .
\end{array}
\label{av_var}
\end{equation}
We restrict our attention to the case of weak disorder, defined by the
condition
\begin{equation}
\sigma^{2} \ll 1 .
\label{weakdis}
\end{equation}
To complete the description of the statistical properties of the
model~(\ref{andmod}), we introduce the  normalised binary correlator
\begin{equation}
\frac{\langle \varepsilon_{n} \varepsilon_{n+l} \rangle}
{\langle \varepsilon_{n}^{2} \rangle} = \chi (l) .
\label{bincor}
\end{equation}
Note that, in the weak-disorder limit, only the binary
correlator~(\ref{bincor}) is required to define the statistical properties
of the disorder (unless one wants to go beyond the second-order
approximation).
We assume that the system is spatially homogeneous in the mean; for this
reason the binary correlator~(\ref{bincor}) depends only on the distance
$l$ between the sites. We further assume disorder to be left-right
symmetric on average so that $\chi(l)$ is an even function of $l$.

We further assume that the binary correlator $\chi(l)$ decreases quickly
beyond a finite length scale $l_{c}$.
This condition was not used in previous second-order analyses of the
Anderson model with weak disorder~\cite{Izr12}.
It is needed here, however, to avoid mathematical inconsistencies in the
application of the special technique which is required to deal with the
compound problem of the band-centre anomaly in the presence of correlated
disorder (see Sec.~\ref{general_formulae}).
We shall consider disorder to be weak enough that condition
\begin{equation}
\sigma^{2} l_{c} \ll 1
\label{shortlc}
\end{equation}
applies. In physical terms, this is equivalent to the assumption that
the correlation length $l_{c}$ is much shorter than the localisation
length.
To deal with the case of long-ranged correlations, we shall derive results
valid for any finite $l_{c}$ and then consider the limit $l_{c} \to \infty$,
as discussed in Sec.~\ref{longrangedcor}.

\subsection{The Hamiltonian map}

The Hamiltonian map approach provides a useful way to study the structure
of the electronic states of the Anderson model~\cite{Izr98}.
The approach is based on the analogy between the quantum model~(\ref{andmod})
and a classical parametric oscillator with Hamiltonian
\begin{equation}
H = \frac{p^{2}}{2m} + \frac{1}{2}m\omega^{2} x^{2} \left[ 1 + \xi(t) \right]
\label{kickos}
\end{equation}
where $\xi(t)$ is a succession of delta kicks of random strengths
\begin{displaymath}
\xi(t) = \sum_{n = -\infty}^{\infty} \xi_{n} \delta \left( t - n T \right).
\end{displaymath}
The integration of the dynamical equations of the oscillator~(\ref{kickos})
over the period $T$ between two kicks leads to the Hamiltonian map
\begin{equation}
\begin{array}{ccl}
x_{n+1} & = & \displaystyle
\left[ \cos \left( \omega T \right) - \omega \xi_{n} \sin \left(
\omega T \right) \right] x_{n} + \frac{1}{m\omega} \sin \left( \omega T
\right) p_{n} \\
p_{n+1} & = & \displaystyle
- m \omega \left[ \sin \left( \omega T \right) + \omega \xi_{n}
\cos \left( \omega T \right) \right] x_{n} +
\cos \left( \omega T \right) p_{n} .
\end{array}
\label{hammap}
\end{equation}
By eliminating the momenta from the map~(\ref{hammap}), one obtains the
equation
\begin{displaymath}
x_{n+1} + x_{n-1} + \omega \xi_{n} \sin \left( \omega T \right)
x_{n} = 2 \cos \left( \omega T \right) x_{n} ,
\end{displaymath}
which coincides with the Schr\"{o}dinger equation~(\ref{andmod})
for the Anderson model provided that
\begin{displaymath}
\begin{array}{ccc}
E = 2 \cos \left( \omega T \right) & \mbox{and} &
\varepsilon_{n} = \omega \xi_{n} \sin \left( \omega T \right) .
\end{array}
\end{displaymath}
It is convenient to write the Hamiltonian map~(\ref{hammap}) in terms of
the action-angle variables $(J_{n},\theta_{n})$, defined by the equations
\begin{displaymath}
\begin{array}{ccl}
x_{n} & = & \displaystyle
\sqrt{\frac{2J_{n}}{m\omega}} \sin \theta_{n} \\
p_{n} & = & \displaystyle
\sqrt{2 m \omega J_{n}} \cos \theta_{n} .
\end{array}
\end{displaymath}
In this way one arrives at the map
\begin{eqnarray}
\begin{array}{c}
\theta_{n+1} \\
\\
\end{array}
&\begin{array}{c} = \\ + \\ \end{array} &
\begin{array}{l} 
\theta_{n} + \omega T + \omega \xi_{n} \sin^{2} \theta_{n}
+ (\omega \xi_{n})^{2} \sin^{3} \theta_{n} \cos \theta_{n} \\
o\left( \sigma^{2} \right) \pmod{2 \pi}
\end{array}
\label{map} \\
J_{n+1} & = & J_{n} \left( 1 - 2 \omega \xi_{n} \sin \theta_{n} \cos \theta_{n}
+ \omega^{2} \xi_{n}^{2} \sin^{2} \theta_{n} \right) .
\nonumber
\end{eqnarray}
In Eq.~(\ref{map}) we have used the Landau symbol $o(\sigma^{2})$ to denote
neglected terms which, in the limit $\sigma \to 0$, vanish faster than
$\sigma^{2}$ (see, e.g.,~\cite{Har60}).
In what follows we shall mostly omit the symbol $o\left( \sigma^{2} \right)$
and tacitly assume that the identities are correct within the limits of the
second-order approximation in the disorder strength.

\subsection{The inverse localisation length}

The inverse localisation length (or Lyapunov exponent) is defined as
\begin{displaymath}
\lambda = \lim_{N \rightarrow \infty} \frac{1}{N} \sum_{n=1}^{N}
\ln \left| \frac{\psi_{n}}{\psi_{n-1}} \right| .
\end{displaymath}
Going to action-angle variables, one can write the Lyapunov exponent as
\begin{equation}
\lambda = \lim_{N \rightarrow \infty} \frac{1}{N} \sum_{n=1}^{N}
\ln \frac{J_{n}}{J_{n-1}} + \lim_{N \rightarrow \infty} \frac{1}{N}
\ln \left| \frac{\sin \theta_{N}}{\sin \theta_{0}} \right| .
\label{lambda0}
\end{equation}
Except that at the band edge (where the angular variable tends to assume
the values $0$ and $\pi$), the second term in the right-hand side (rhs)
of Eq.~(\ref{lambda0}) vanishes; one is therefore left with
\begin{equation}
\lambda = \lim_{N \rightarrow \infty} \frac{1}{N} \sum_{n=1}^{N}
\ln \frac{J_{n}}{J_{n-1}} =
\langle \ln \left( 1 - 2 \omega \xi_{n} \sin \theta_{n} \cos \theta_{n}
+ \omega^{2} \xi_{n}^{2} \sin^{2} \theta_{n} \right) \rangle .
\label{lambda1}
\end{equation}
For weak disorder it is possible to expand the logarithm in the
rhs of Eq.~(\ref{lambda1}) and write
\begin{equation}
\lambda = \frac{\omega^{2}}{8} \langle \xi_{n}^{2} \rangle
\left[ 1 - 2 \langle \cos \left( 2 \theta_{n} \right) \rangle +
\langle \cos \left( 4 \theta_{n} \right) \rangle \right] - \frac{\omega}{2}
\langle \xi_{n} \sin \left( 2 \theta_{n} \right) \rangle.
\label{lyap1}
\end{equation}

The noise-angle correlator
$\langle  \xi_{n} \sin \left( 2 \theta_{n} \right) \rangle$ vanishes if
the random site energies are independent, but is different from zero for
correlated disorder.
It can be evaluated with the method presented in Ref.~\cite{Izr99};
substituting the result in Eq.~(\ref{lyap1}) one obtains
\begin{equation}
\begin{array}{ccl}
\lambda & = & \displaystyle
\frac{\sigma^{2}}{8 \sin^{2}(\omega T)}
\left\{ \left[ 1 - 2 \langle \cos \left( 2 \theta_{n} \right) \rangle +
\langle \cos \left( 4 \theta_{n} \right) \rangle \right] W(\omega T) \right. \\
& + & \displaystyle \left.
\left[ 2 \langle \sin \left( 2 \theta_{n} \right) \rangle -
\langle \sin \left( 4 \theta_{n} \right) \rangle \right] Y(\omega T) 
\right\}, \\
\end{array}
\label{lyap2}
\end{equation}
where
\begin{equation}
W(x) = 1 + 2 \sum_{l=1}^{\infty} \chi(l) \cos \left( 2 x l \right)
\label{power_spectrum}
\end{equation}
is the power spectrum of the disorder and
\begin{equation}
Y(x) = 2 \sum_{l=1}^{\infty} \chi(l) \sin \left( 2 x l \right)
\label{sine_transform}
\end{equation}
is the sine transform of the binary correlator~(\ref{bincor}).

To evaluate the averages of the trigonometric functions in the rhs of
Eq.~(\ref{lyap2}), it is necessary to determine the distribution of the
angle variable $\theta$.
If $\omega T$ does not lie too close to a value of the form $\pi p/q$ (with
$p$ and $q$ integer numbers), one can see from the map~(\ref{map}) that
the angular variable quickly sweeps the $[0,2\pi]$ interval, thus producing
a uniform invariant distribution,
\begin{displaymath}
\rho(\theta) = \frac{1}{2 \pi} .
\end{displaymath}
If this distribution is used to compute the averages of the trigonometric
functions in Eq.~(\ref{lyap2}), one immediately obtains the standard
formula derived by Izrailev and Krokhin in Ref.~\cite{Izr99},
\begin{equation}
\lambda_{\rm IK} = \frac{\sigma^{2}}{8 \sin^{2} \left(\omega T \right)}
W(\omega T) .
\label{nonres_lyap}
\end{equation}
When $\omega T$ is a rational multiple of $\pi$, however, the
map~(\ref{map}) has almost periodic orbits whose existence manifests
itself in the form of a periodic modulation of $\rho(\theta)$.
This is what happens at the band centre, which corresponds to the value
$\omega T = \pi/2$.
In order to obtain the correct localisation length close to the band
centre, one must therefore determine the invariant distribution
$\rho (\theta)$ for the angular map~(\ref{map}) with $\omega T \simeq \pi/2$.

\subsection{The angle map in a neighbourhood of the band centre}

We consider the case in which
\begin{equation}
\begin{array}{ccc}
\displaystyle
\omega T = \frac{\pi}{2} + \delta & \mbox{ with } & \delta \to 0 .
\end{array}
\label{nbc}
\end{equation}
The corresponding energies lie close to the band centre,
\begin{displaymath}
E = -2 \sin \delta \simeq -2 \delta.
\end{displaymath}
When the parameter $\omega T$ takes the value~(\ref{nbc}), the angular
map~(\ref{map}) has almost-periodic orbits of period 4 and the
difference $\theta_{n+4} - \theta_{n}$ is small. Iterating four times the
angular map~(\ref{map}) leads to
\begin{equation}
\begin{array}{ccl}
\theta_{n+4} & = & \displaystyle
\theta_{n} + 4 \delta + \frac{\varepsilon_{n} + \varepsilon_{n+2}}{2}
\left[ 1 - \cos \left( 2 \theta_{n} \right) \right] +
\frac{\varepsilon_{n+1} + \varepsilon_{n+3}}{2} \left[ 1 +
\cos  \left( 2 \theta_{n} \right) \right]
\\
& - & \displaystyle
\frac{\sigma^{2}}{2} \left[ \chi_{1} + \chi(3) \right]
\sin \left( 2 \theta_{n} \right) +
\frac{\sigma^{2}}{4}\left[ 2 - 3 \chi(1) + 2 \chi(2) - \chi(3) \right]
 \sin \left( 4 \theta_{n} \right)  \\
& & \pmod{2 \pi} .\\
\end{array}
\label{fourth}
\end{equation}
In the rhs of Eq.~(\ref{fourth}) we have neglected terms of order
$O \left( \sigma \delta \right)$ and $o \left( \sigma^{2} \right)$.
In order to determine the invariant measure $\rho(\theta)$, it is
useful to go to the continuum limit and replace the random map~(\ref{fourth})
with an appropriate stochastic differential equation. We devote the next
section to the discussion of how this can be done.

\section{The continuum limit}
\label{rm_sde}

We need a systematic recipe to associate a stochastic differential
equation to a random map of the form~(\ref{fourth}).
Devising such a method is equivalent to finding an integration scheme
for stochastic differential equations with coloured noise.
Although there is an enormous literature on the numerical integration
of stochastic differential equations with {\em white} noise (see~\cite{Klo92}
and references therein), much less is known on how to deal with differential
equations with {\em coloured} noise.
Equations with coloured noise are often reduced to systems of coupled
equations with white noise with the standard trick of representing the
coloured noise as the solution of an extra stochastic differential
equation (see, e.g.~\cite{Fox77,Van07}).
In essence, this method replaces an equation of the form
\begin{displaymath}
\dot{x} = a(x) + b(x) \eta(t) ,
\end{displaymath}
where $\eta(t)$ is a coloured noise, with the pair of coupled equations
\begin{displaymath}
\begin{array}{ccl}
\dot{x} & = & a(x) + b(x) \eta(t) \\
\dot{\eta} & = & -\gamma \eta(t) + L(t) \\
 \end{array}
\end{displaymath}
with $L(t)$ being a white noise.
Numerical integration schemes have been created for equations of this
kind (see, e.g.,~\cite{Man89}). This approach, however, can be applied
only if the coloured noise is exponentially correlated, and we would like
to avoid such a constraint.
For this reason, we have derived a new integration scheme, which does not
suffer from the same limitations.

Let us consider a stochastic differential equation of the form
\begin{equation}
\dot{x} = a(x,t) + \sum_{i=1}^{N} b^{(i)}(x,t) \zeta_{i}(t)
\label{proto_sde}
\end{equation}
where $a(x,t)$ and the $b^{(i)}(x,t)$ are deterministic functions
(with the functions $b^{(i)}(x,t)$ being differentiable), while
the $\zeta_{i}(t)$ are stochastic processes with zero averages,
\begin{equation}
\langle \zeta_{i}(t) \rangle = 0
\label{av_xi}
\end{equation}
and correlation matrix of the form
\begin{equation}
\langle \zeta_{i}(t) \zeta_{j}(t+\tau) \rangle = \chi_{ij}(\tau) .
\label{cor_xi}
\end{equation}
The matrix elements $\chi_{ij}(\tau)$ in Eq.~(\ref{cor_xi}) are assumed
to be even and decreasing functions of the time difference $\tau$.
Since we are interested in the case of weak noise, we will not specify
the statistical properties of the $\zeta_{i}(t)$ processes in further detail.
If the $\zeta_{i}(t)$ are independent white noises, i.e., Gaussian
``processes'' with correlation functions
\begin{displaymath}
\langle \zeta_{i}(t) \zeta_{j}(t+\tau) \rangle = \delta_{ij} \delta(\tau) ,
\end{displaymath}
we can interpret Eq.~(\ref{proto_sde}) as an informal way to write the
Stratonovich stochastic differential equation
\begin{displaymath}
\mathrm{d}x = a(x,t) \mathrm{d}t + \sum_{i=1}^{N} b^{(i)}(x,t) \circ
\mathrm{d} W_{i} .
\end{displaymath}

Following~\cite{Man02}, we integrate the stochastic equation~(\ref{proto_sde})
over a short time interval $[t,t+\epsilon]$ and we obtain
\begin{displaymath}
x(t + \epsilon) = x(t) + \int_{t}^{t + \epsilon} \left[ a(x(\tau),\tau) +
\sum_{i=1}^{N} b^{(i)}(x(\tau),\tau) \zeta_{i}(\tau) \right] \mathrm{d} \tau .
\end{displaymath}
A recursive application of this identity gives
\begin{displaymath}
\begin{array}{ccl}
x(t + \epsilon) & = & \displaystyle
x(t) + a(x(t),t) \epsilon +
\sum_{i} b^{(i)}(x(t),t) \int_{t}^{t + \epsilon} \mathrm{d} \tau \; \zeta_{i}(\tau) \\
& + & \displaystyle
\sum_{ij} \frac{\partial b^{(i)}}{\partial x}(x(t),t) b^{(j)}(x(t),t)
\int_{t}^{t + \epsilon} \mathrm{d} \tau_{1}
\int_{t}^{t + \epsilon} \mathrm{d} \tau_{2} \; 
\zeta_{i}(\tau_{1}) \zeta_{j}(\tau_{2}) + \ldots \\
\end{array}
\end{displaymath}
Neglecting the fluctuations of the noisy quadratic term results in the
following integration scheme
\begin{equation}
\begin{array}{ccl}
x(t + \epsilon) & = & \displaystyle
x(t) + a(x(t),t) \epsilon + \sum_{ij} I_{ij}
\frac{\partial b^{(i)}}{\partial x}(x(t),t) b^{(j)}(x(t),t) \\
& + & \displaystyle
\sum_{i} b^{(i)}(x(t),t) \int_{t}^{t+\epsilon} \zeta_{i}(\tau) \mathrm{d} \tau
+\ldots
\end{array}
\label{scheme1}
\end{equation}
where the symbol $I_{ij}$ stands for the integral
\begin{equation}
I_{ij} = \frac{1}{2} \int_{-\epsilon}^{\epsilon} \left( \epsilon - |\tau| \right)
\chi_{ij}(-|\tau|) \mathrm{d} \tau .
\label{stoc_int}
\end{equation}
It is now convenient to consider the discrete times $t_{n} = n \epsilon$ and
to introduce the compact notations
\begin{equation}
\begin{array}{ccc}
x_{n} = x(t_{n}), & & a_{n} = a(x_{n},t_{n}), \\
b^{(i)}_{n} = b^{(i)}(x_{n},t_{n}), & &
\displaystyle \frac{\partial b^{(i)}_{n}}{\partial x} =
\frac{\partial b^{(i)}}{\partial x}(x_{n},t_{n}) . \\
\end{array}
\label{compact_notation}
\end{equation}
We also define the new random variables
\begin{equation}
Z_{n}^{(i)} = \int_{t_{n}}^{t_{n+1}} \zeta_{i}(\tau) \mathrm{d} \tau .
\label{bigxi}
\end{equation}
The statistical properties of the noises $\zeta_{i}(t)$ define the
corresponding properties of the random variables~(\ref{bigxi}). In particular,
Eqs.~(\ref{av_xi}) and~(\ref{cor_xi}) imply that
\begin{displaymath}
\langle Z_{n}^{(i)} \rangle = 0
\end{displaymath}
and
\begin{displaymath}
\langle Z_{n}^{(i)}\; Z_{n+k}^{(j)} \rangle =
\int_{-\epsilon}^{\epsilon} \left( \epsilon - |\tau| \right)
\chi_{ij}(\tau + k \epsilon) \mathrm{d} \tau .
\end{displaymath}
We can now write Eq.~(\ref{scheme1}) in the form of a map
\begin{equation}
x_{n+1} = x_{n} \epsilon + \sum_{ij} \frac{\partial b^{(i)}_{n}}{\partial x}
b^{(j)}_{n} I_{ij} + \sum_{i} b^{(i)}_{n} \; Z^{(i)}_{n} + \ldots
\label{scheme2}
\end{equation}
The map~(\ref{scheme2}) represents an integration scheme for the stochastic
differential equation~(\ref{proto_sde}).

We now focus our attention to the case in which the binary
correlator~(\ref{cor_xi}) has the form
\begin{equation}
\langle \zeta_{i}(t) \zeta_{j}(t+\tau) \rangle 
= \sum_{k=-\infty}^{\infty} \mathrm{X}_{ij}(4k) \delta(\tau - k\epsilon) ,
\label{cor_xi2}
\end{equation}
with $\mathrm{X}_{ij}(4k)$ being a decreasing function of the argument $4k$.
In this case the correlation function of the random variables~(\ref{bigxi})
becomes
\begin{equation}
\langle Z_{n}^{(i)}\; Z_{n+k}^{(j)} \rangle =
\epsilon \mathrm{X}_{ij}(4k) .
\label{cor_bigxi}
\end{equation}
Eq.~(\ref{cor_xi2}) also allows one to write the integrals~(\ref{stoc_int}) in
the form
\begin{displaymath}
I_{ij} = \frac{\epsilon}{2} \mathrm{X}_{ij}(0) .
\end{displaymath}
Substituting this identity in the map~(\ref{scheme2}) and setting
$\epsilon = 1$, we finally obtain
\begin{equation}
x_{n+1} = x_{n} + \frac{1}{2} \sum_{ij} \mathrm{X}_{ij}(0)
\frac{\partial b^{(i)}_{n}}{\partial x} b^{(j)}_{n}  +
\sum_{i} b^{(i)}_{n} \; Z^{(i)}_{n} .
\label{scheme3}
\end{equation}
The map~(\ref{scheme3}) contains the random variables $Z^{(i)}_{n}$ with zero
averages and correlation function defined by Eq.~(\ref{cor_bigxi}).
Our derivation shows that the map~(\ref{scheme3}) represents an integration
scheme for the stochastic differential equation~(\ref{proto_sde}), which
contains noises $\zeta_{i}(t)$ having vanishing averages and correlation
function~(\ref{cor_xi2}).

We can now address the inverse problem, i.e., how to associate a stochastic
differential equation to a given a random map.
Let us consider the map
\begin{equation}
x_{n+1} = x_{n} + a_{n} + \sum_{i} b_{n}^{(i)} \; Z^{(i)}_{n} ,
\label{random_map}
\end{equation}
where $a_{n}$ and $b^{(i)}_{n}$ are short-hand notations defined
by Eq.~(\ref{compact_notation}) and the symbols $Z^{(i)}_{n}$ represent
random variables with zero average and binary correlator of the
form~(\ref{cor_bigxi}).
The correspondence between the stochastic differential
equation~(\ref{proto_sde}) and the map~(\ref{scheme3}) implies that the
random map~(\ref{random_map}) can be read as an integration scheme for
the stochastic differential equation
\begin{equation}
\begin{array}{ccl}
\dot{x} & = & \displaystyle
a(x,t) - \frac{1}{2} \sum_{ij} \mathrm{X}_{ij}(0)
\frac{\partial b^{(i)}}{\partial x}(x,t) b^{(j)}(x,t) \\
& + & \displaystyle
\sum_{i} b^{(i)}(x,t) \zeta_{i}(t) , \\
\end{array}
\label{conti_lim}
\end{equation}
where the stochastic processes $\zeta_{i}(t)$ have zero averages and
correlation function given by Eq.~(\ref{cor_xi2}).

To test the correspondence between the map~(\ref{random_map}) and the
stochastic differential equation~(\ref{conti_lim}), we can apply it to
the case of the random map~(\ref{map}) for the angle variable.
In this case, the continuum limit is a differential equation of the form
\begin{equation}
\dot{\theta} = \omega T + \zeta(t) \sin^{2} \theta 
\label{old_cl}
\end{equation}
where $\zeta(t)$ is a noise with zero average and correlation function
\begin{equation}
\langle \zeta(t) \zeta(t+\tau) \rangle = \omega^{2} \sum_{k=-\infty}^{\infty}
\langle \xi_{n} \xi_{n+k} \rangle \delta \left( \tau - kT \right) .
\label{old_corr}
\end{equation}
Eqs.~(\ref{old_cl}) and~(\ref{old_corr}) define a dynamical system
which was shown to be the continuous counterpart of the discrete
Anderson model~(\ref{andmod}) for non-resonant values of the
energy~\cite{Tes01}.
In Ref.~\cite{Tes01}, however, the correspondence between the two models
was established only retrospectively, by computing separately the Lyapunov
exponents of both systems and observing that they coincide.
The integration scheme derived in this section, on the other hand, has enabled
us to predict {\em a priori} that the dynamical system~(\ref{old_cl})
is the continuous analogue of the discrete Anderson model~(\ref{andmod}).
This is of crucial importance in the present case, where the continuum
limit is instrumental in deriving an expression for the localisation
length at the band centre.

\section{Invariant measure and localisation length in a neighbourhood of the
band centre}
\label{general_formulae}

\subsection{The stochastic differential equation}

Having established a correspondence between a generic random map of
the form~(\ref{random_map}) and the stochastic differential
equation~(\ref{conti_lim}), we can now consider the continuum limit of
the specific map~(\ref{fourth}).
As a first step, we can write the map~(\ref{fourth}) in the form
\begin{equation}
\begin{array}{ccl}
\theta_{n+4} & = & \displaystyle
\theta_{n} + 4 \delta - \frac{\sigma^{2}}{2} \left[ \chi(1)
+ \chi(3)\right] \sin \left( 2 \theta_{n} \right) \\
& - & \displaystyle
\frac{\sigma^{2}}{4} \left[ 2 - 3 \chi(1) +2 \chi(2) - \chi(3) \right]
\sin \left( 4 \theta_{n} \right) \\
& + & \displaystyle
\left[ 1 - \cos \left( 2 \theta_{n} \right) \right] Z^{(1)}_{n} +
\left[ 1 + \cos \left( 2 \theta_{n} \right) \right] Z^{(2)}_{n} \\
\end{array}
\label{fourth_bis}
\end{equation}
with
\begin{equation}
\begin{array}{ccc}
\displaystyle
Z^{(1)}_{n} = \frac{\varepsilon_{n} + \varepsilon_{n+2}}{2} &
\mbox{ and } & \displaystyle
Z^{(2)}_{n} = \frac{\varepsilon_{n+1} + \varepsilon_{n+3}}{2} .
\end{array}
\label{bigxi2}
\end{equation}
The random variables~(\ref{bigxi2}) have zero average and Eq.~(\ref{bincor})
implies that their correlation matrix has elements
\begin{displaymath}
\begin{array}{l}
\displaystyle
\langle Z^{(1)}_{n} \; Z^{(1)}_{n+4k} \rangle =
\langle Z^{(2)}_{n} \; Z^{(2)}_{n+4k} \rangle =
\frac{\sigma^{2}}{4} \left[ 2 \chi(4k) + \chi(4k+2) + \chi(4k-2) \right] \\
\displaystyle
\langle Z^{(1)}_{n} \; Z^{(2)}_{n+4k} \rangle =
\frac{\sigma^{2}}{4} \left[ 2 \chi(4k+1) + \chi(4k+3) + \chi(4k-1) \right] \\
\displaystyle
\langle Z^{(2)}_{n} \; Z^{(1)}_{n+4k} \rangle =
\frac{\sigma^{2}}{4} \left[ 2 \chi(4k-1) + \chi(4k+1) + \chi(4k-3) \right] . \\
\end{array}
\end{displaymath}

Following the approach discussed in Sec.~\ref{rm_sde}, we can interpret the
map~(\ref{fourth_bis}) as the integration scheme with time step
\begin{displaymath}
\epsilon = 4T = 1
\end{displaymath}
of a stochastic differential equation of the form
\begin{equation}
\dot{\theta} = F^{(0)}(\theta) + F^{(1)}(\theta,t),
\label{sde}
\end{equation}
where $F^{(0)}(\theta)$ is the deterministic function
\begin{displaymath}
F^{(0)}(\theta) = 4 \delta - \frac{\sigma^{2}}{2}
\left[ \chi(1) + \chi(3) \right] \sin \left( 2 \theta \right) ,
\end{displaymath}
while the stochastic part is given by
\begin{equation}
F^{(1)}(\theta,t) = \zeta_{1}(t) \left[ 1 - \cos \left( 2 \theta \right) \right]
+ \zeta_{2}(t) \left[ 1 + \cos \left( 2 \theta \right) \right] .
\label{stoc_part}
\end{equation}
In Eq.~(\ref{stoc_part}), $\zeta_{1}(t)$ and $\zeta_{2}(t)$ are two
cross-correlated coloured noises with zero average and binary correlators
\begin{equation}
\langle \zeta_{i}(t) \zeta_{j}(t+\tau) \rangle =
\sum_{k=-\infty}^{\infty} \langle Z^{(i)}_{n} \; Z^{(j)}_{n+4k} \rangle
\delta \left( \tau - k \right)
\label{xi_corr}
\end{equation}
with $i,j=1,2$.
Note that the condition that the binary correlator~(\ref{bincor}) decays
quickly over distances larger than $l_{c}$ implies that the
correlators~(\ref{xi_corr}) also vanish over time scales
$\tau \gg \tau_{c} \sim l_{c}$.

We remark that, in the limit case of uncorrelated disorder, Eq.~(\ref{sde})
reduces to
\begin{displaymath}
\mathrm{d} \theta = 4 \delta \mathrm{d} t + \sqrt{\frac{\sigma^{2}}{2}}
\left[ 1 - \cos \left( 2 \theta \right) \right] \circ \mathrm{d} W_{1} +
\sqrt{\frac{\sigma^{2}}{2}}
\left[ 1 + \cos \left( 2 \theta \right) \right] \circ \mathrm{d} W_{2} ,
\end{displaymath}
which is the Stratonovich form of the It\^{o} equation used in~\cite{Tes12}
to analyse the band-centre anomaly in the case of uncorrelated disorder.

\subsection{Associated Fokker-Planck equation}

To proceed further, we make use of assumption~(\ref{shortlc}), which we
can now write in the equivalent form
\begin{equation}
\tau_{c} \sigma^{2} \ll 1 .
\label{shorttc1}
\end{equation}
As shown by Van Kampen~\cite{Van07}, if condition~(\ref{shorttc1}) holds
one can associate a Fokker-Planck equation to the stochastic differential
equation~(\ref{sde}). The statistical properties of the solution $\theta(t)$
of Eq.~(\ref{sde}) can then be described in terms of a probability
$P(\theta,t)$ which obeys the associated Fokker-Planck equation
\begin{equation}
\frac{\partial P}{\partial t}(\theta,t) = -\frac{\partial}{\partial \theta}
\left[ C_{1}(\theta) P(\theta,t) \right] +
\frac{\partial^{2}}{\partial \theta^{2}} \left[ C_{2}(\theta) P(\theta,t) \right].
\label{fp1}
\end{equation}
The drift and diffusion coefficients in Eq.~(\ref{fp1}) are given by
\begin{equation}
C_{1}(\theta) = F^{(0)}(\theta) + \int_{0}^{\infty} \langle
\frac{\partial F^{(1)}}{\partial \theta}(\theta, t)
F^{(1)}(\theta^{-\tau}, t - \tau) \rangle \frac{d \theta}{d \theta^{-\tau}}
\mathrm{d} \tau 
\label{driftcoef}
\end{equation}
and
\begin{equation}
C_{2}(\theta) = \int_{0}^{\infty} \langle F^{(1)}(\theta, t)
F^{(1)}(\theta^{-\tau},t-\tau)  \rangle \mathrm{d} \tau .
\label{difcoef}
\end{equation}
In Eqs.~(\ref{driftcoef}) and~(\ref{difcoef}), the symbol $\theta^{t}$
represents the solution of the ordinary differential equation
\begin{displaymath}
\dot{\theta}^{t} = F^{(0)}(\theta^{t})
\end{displaymath}
with initial condition $\theta^{0} = \theta$.

To simplify the mathematical expressions, in what follows we assume that
$\tau_{c}$ satisfies not only condition~(\ref{shorttc1}) but also the
additional condition
\begin{equation}
\tau_{c} \delta \ll 1 .
\label{shorttc2}
\end{equation}
The present approach can be applied even if condition~(\ref{shorttc2})
does not hold; assuming that it does, however, is mathematically
convenient because the combination of conditions~(\ref{shorttc1})
and~(\ref{shorttc2}) ensures that $\theta^{t} \simeq \theta$ over time
scales $|t| \lesssim \tau_{c}$ and therefore one can replace $\theta^{-\tau}$
with $\theta$ in the integral expressions
of the coefficients~(\ref{driftcoef}) and~(\ref{difcoef}). This leads
to a significant simplification of the mathematical expressions.
From a physical point of view, condition~(\ref{shorttc2}) is not very
restrictive, because we are interested in the neighbourhood of the
band centre, which corresponds to the limit $\delta \to 0$.

Carrying out the calculations, one obtains that the Fokker-Planck
equation~(\ref{fp1}) takes the specific form
\begin{equation}
\begin{array}{ccl}
\displaystyle
\frac{\partial P}{\partial t}(\theta,t) & = & \displaystyle
\frac{\sigma^{2}}{4} \frac{\partial}{\partial \theta} \Bigg\{
\left[ 8 \varkappa - 2 W \left( \frac{\pi}{2} \right) \sin (4 \theta) \right]
P(\theta,t) \\
& + & \displaystyle
\left[ 2 W \left( 0 \right) + W \left( \frac{\pi}{2} \right)
\left[ 1 + \cos ( 4 \theta) \right] \right]
\frac{\partial P}{\partial \theta} (\theta,t) \Bigg\} ,
\end{array}
\label{fp2}
\end{equation}
where $W$ is the power spectrum~(\ref{power_spectrum}) and we have
introduced the parameter
\begin{equation}
\varkappa = -\frac{2 \delta}{\sigma^{2}} .
\label{kappa}
\end{equation}

\subsection{The invariant measure}

We are interested in the stationary solution of the Fokker-Planck
equation~(\ref{fp2}), i.e., in the function
\begin{equation}
\rho(\theta) = \lim_{t \to \infty} P(\theta,t),
\label{invmeas}
\end{equation}
which represents the invariant measure of the map~(\ref{fourth}).
The distribution~(\ref{invmeas}) satisfies the first-order differential
equation
\begin{equation}
\frac{d \rho}{d \theta}(\theta) = A(\theta) \rho(\theta) +
C B(\theta)
\label{ode}
\end{equation}
where $C$ is an integration constant while $A(\theta)$ and $B(\theta)$ are
the functions
\begin{displaymath}
A(\theta) = \frac{2 W(\pi/2) \sin (4 \theta) - 8 \varkappa}{2 W(0) +
W(\pi/2) \left[ 1 + \cos (4 \theta) \right]}
\end{displaymath}
and
\begin{displaymath}
B(\theta) = \frac{1}{2 W(0) + W(\pi/2) \left[ 1 + \cos (4 \theta) \right]} .
\end{displaymath}
The general solution of Eq.~(\ref{ode}) is
\begin{equation}
\begin{array}{ccl}
\rho(\theta) & = & \displaystyle
\frac{e^{-8 \varkappa F(\theta)}}
{\sqrt{2 W(0) + W(\pi/2) \left[ 1 + \cos (4 \theta) \right]}}
\left\{ \sqrt{2 \left[ W(0) + W(\pi/2) \right]} \rho(0) \right. \\
& + & \displaystyle \left.
C \int_{0}^{\theta} \frac{e^{8 \varkappa F(\phi)}}
{\sqrt{2 W(0) + W(\pi/2) \left[ 1 + \cos (4 \theta) \right]}}
\mathrm{d} \phi \right\} \\
\end{array}
\label{gensol}
\end{equation}
The function $F(\theta)$ in Eq.~(\ref{gensol}) is defined by the integral
representation
\begin{displaymath}
F(\theta) = \int_{0}^{\theta}
\frac{1}{2 W(0) + W(\pi/2) \left[ 1 + \cos (4 \phi) \right]} \;
\mathrm{d} \phi .
\end{displaymath}
Carrying out the integration one obtains the explicit expression
\begin{equation}
F(\theta) = \frac{1}{4 \sqrt{W(0) \left[ W(0) + W(\pi/2) \right]}}
\left\{ \arctan \left[ \sqrt{\frac{W(0)}{W(0) + W(\pi/2)}} \tan \left(
2 \theta \right) \right] + \pi n \right\}
\label{f}
\end{equation}
for $(2n-1) \pi/4 \leq \theta \leq (2n+1) \pi/4$ and $n \in \mathbf{Z}$.
In Eq.~(\ref{f}), $\arctan(x)$ is a function with principal values in the
interval $[-\pi/2,\pi/2]$. Note that
\begin{displaymath}
F(2 \pi) = \frac{\pi}{\sqrt{W(0) \left[ W(0) + W(\pi/2) \right]}}.
\end{displaymath}

The constant $C$ in Eq.~(\ref{gensol}) can be expressed in terms of
$\rho(0)$ with the help of the periodicity condition $\rho(2\pi) = \rho(0)$.
The normalisation condition then determines the remaining integration
constant $\rho(0)$. In this way one arrives at the desired invariant
measure
\begin{equation}
\begin{array}{ccl}
\rho(\theta, \varkappa) & = & \displaystyle
\frac{e^{-8 \varkappa F(\theta)}}
{\sqrt{2 W(0) + W(\frac{\pi}{2}) \left[ 1 + \cos (4 \theta) \right]}}
\left\{ \int_{\theta}^{2\pi} \frac{e^{8 \varkappa F(\phi)}}
{\sqrt{2 W(0) + W(\frac{\pi}{2}) \left[ 1 + \cos (4 \phi) \right]}}
\mathrm{d} \phi \right. \\
& + & \displaystyle \left.
e^{8 \pi F(2 \pi)} \int_{0}^{\theta} \frac{e^{8 \varkappa F(\phi)}}
{\sqrt{2 W(0) + W(\frac{\pi}{2}) \left[ 1 + \cos (4 \phi) \right]}}
\mathrm{d} \phi \right\} \frac{1}{N(\varkappa)} , \\
\end{array}
\label{genrho}
\end{equation}
where $N(\varkappa)$ is a normalisation constant equal to
\begin{displaymath}
\begin{array}{lcl}
N(\varkappa) & = & \displaystyle
\int_{0}^{2\pi} \frac{e^{-8 \varkappa F(\theta)}}
{\sqrt{2 W(0) + W(\frac{\pi}{2}) \left[ 1 + \cos (4 \theta) \right]}} 
\left\{ \int_{\theta}^{2\pi} \frac{e^{8 \varkappa F(\phi)}}
{\sqrt{2 W(0) + W(\frac{\pi}{2}) \left[ 1 + \cos (4 \phi) \right]}}
\mathrm{d} \phi \right. \\
& + & \displaystyle \left.
e^{8 \pi F(2 \pi)} \int_{0}^{\theta} \frac{e^{8 \varkappa F(\phi)}}
{\sqrt{2 W(0) + W(\frac{\pi}{2}) \left[ 1 + \cos (4 \phi) \right]}}
\mathrm{d} \phi \right\} \mathrm{d} \theta .\\
\end{array}
\end{displaymath}
In Eq.~(\ref{genrho}) we have written $\rho$ as a function of two arguments,
$\theta$ and $\varkappa$, to stress that the invariant distribution of the
angular variable $\theta$ depends on the energy parameter~(\ref{kappa}).

A key feature of the invariant distribution~(\ref{genrho}) is that it
is $\pi/2$-periodic:
\begin{equation}
\rho \left(\theta + \frac{\pi}{2},\varkappa \right) =
\rho \left(\theta,\varkappa \right) .
\label{pi2periodicity}
\end{equation}
The demonstration of Eq.~(\ref{pi2periodicity}) can be carried out along
the same lines followed in the case of uncorrelated disorder~\cite{Tes12}.
At first sight, Eq.~(\ref{pi2periodicity}) may seem surprising because
the deterministic part of the map~(\ref{fourth_bis}) has stable
zero-velocity points for $\theta = 0$ and $\theta = \pm \pi$.
One could therefore expect probability peaks to appear at these points and
the distribution $\rho(\theta,\varkappa)$ to be $\pi$-periodic.
However, a closer examination of the map~(\ref{fourth_bis})
reveals that the noisy terms are largest exactly for $\theta = 0$ and
$\theta = \pm \pi$. The noise, therefore, scatters the angle variable
away from the stable zero-velocity points.
Since the deterministic and the noisy terms in the map~(\ref{fourth_bis})
carry the same weight at the band centre, the period of the invariant
distribution ends up being $\pi/2$ rather than $\pi$.

To conclude our analysis of the invariant distribution~(\ref{genrho}), we
can consider its limit forms for $\varkappa = 0$ and $\varkappa \to \infty$.
At the exact band centre ($\varkappa = 0$), Eq.~(\ref{genrho}) reduces to
the significantly simpler form
\begin{equation}
\rho(\theta, \varkappa=0) = 
\frac{1}{\displaystyle 2 \mathbf{K}
\left( \sqrt{\frac{W(\pi/2)}{W(0)+W(\pi/2)}} \right)}
\frac{1}{\displaystyle \sqrt{4 - \frac{2 W(\pi/2)}{W(0) + W(\pi/2)}
\left[ 1 - \cos (4 \theta) \right]}} ,
\label{bcrho}
\end{equation}
where the symbol $\mathbf{K}(k)$ represents the complete elliptic integral
of the first kind.
In the opposite limit, i.e., for $\varkappa \to \infty$, the
asymptotic expansion of the general formula~(\ref{genrho}) gives
\begin{equation}
\rho(\theta,\varkappa) = \frac{1}{2\pi} \left[ 1 + \frac{1}{4 \varkappa}
W\left( \frac{\pi}{2} \right) \sin \left( 4 \theta \right) + \ldots \right].
\label{asymptotic_kappa}
\end{equation}
Eq.~(\ref{asymptotic_kappa}) shows that, when the energy moves away from the
band centre, the difference between the invariant distribution and its
flat limit form falls off as the inverse of the energy.

\subsection{The localisation length}

Having determined the invariant measure~(\ref{genrho}), we can now turn our
attention to the localisation length. Since $\rho(\theta,\varkappa)$ is
$\pi/2$-periodic, one has
\begin{displaymath}
\langle \cos (2 \theta) \rangle = \langle \sin (2 \theta) \rangle = 0.
\end{displaymath}
Therefore, in a neighbourhood of the band centre, Eq.~(\ref{lyap2}) reduces
to
\begin{equation}
\lambda = \frac{\sigma^{2}}{8 \cos^{2}(\delta)} \left\{ \left[ 1 +
\langle \cos ( 4 \theta ) \rangle \right]
W \left( \frac{\pi}{2} + \delta \right) - \langle \sin ( 4 \theta ) \rangle
Y \left( \frac{\pi}{2} + \delta \right) \right\} ,
\label{lyap3}
\end{equation}
with the functions $W$ and $Y$ defined by Eqs.~(\ref{power_spectrum})
and~(\ref{sine_transform}).
Taking into account that $Y(\pi/2) = 0$ and neglecting terms of order
$O(\sigma^{2} \delta)$, one can write
Eq.~(\ref{lyap3}) as
\begin{equation}
\lambda = \frac{\sigma^{2}}{8}
\left[ 1 + \langle \cos \left( 4 \theta_{n} \right) \rangle \right]
W \left( \frac{\pi}{2} \right) ,
\label{lyap}
\end{equation}
where the average of the trigonometric function must be computed using
the invariant measure~(\ref{genrho}).

Eq.~(\ref{lyap}) represents the expression for the inverse localisation
length in a neighbourhood of the band centre. It should be compared with
the formula~(\ref{nonres_lyap}) for non-resonant energies which, close to the
band centre, reduces to
\begin{equation}
\lambda_{\rm IK} = \frac{\sigma^{2}}{8} W \left( \frac{\pi}{2} \right) .
\label{nonres_lyap_bc}
\end{equation}
As can be seen from Eqs.~(\ref{lyap}) and~(\ref{nonres_lyap_bc}), the
modulation of the invariant distribution produces an anomaly in the
inverse localisation length because the term $\langle \cos (4\theta) \rangle$
does not vanish.
If the inverse localisation length~(\ref{lyap}) is compared with the
expression for uncorrelated disorder, it becomes obvious that spatial
correlations have a twofold effect on the localisation length.
On the one hand, the correlations manifest themselves via the power-spectrum
factor $W(\pi/2)$, which is present also in the standard
formula~(\ref{nonres_lyap_bc}).
On the other hand, the correlations modify the invariant distribution
$\rho(\theta,\varkappa)$ and therefore affect the value of the averaged
cosine in Eq.~(\ref{lyap}).

An explicit evaluation of the $\langle \cos (4\theta) \rangle$
is not possible in the general case, but it can be obtained at the exact
band centre, where the invariant distribution takes the simpler
form~(\ref{bcrho}).
In this case the inverse localisation length~(\ref{lyap}) can be written as
\begin{equation}
\lambda (\varkappa = 0) = \frac{\sigma^{2}}{4} \left\{ \left[ W(0) +
W \left( \frac{\pi}{2} \right) \right] \frac{\displaystyle \mathbf{E}
\left( \sqrt{\frac{W(\pi/2)}{W(0) + W(\pi/2)}} \right)}
{\displaystyle \mathbf{K}
\left( \sqrt{\frac{W(\pi/2)}{W(0) + W(\pi/2)}} \right)} - W(0) \right\}
\label{bclyap}
\end{equation}
where $\mathbf{K}(k)$ and $\mathbf{E}(k)$ represent the complete
elliptic integrals of the first and second kind.

Eqs.~(\ref{genrho}) and~(\ref{lyap}) constitute the central results of this
paper. They provide general formulae for the invariant measure of the angular
variable and for the localisation length of the Anderson model~(\ref{andmod}).
To understand how the correlations of the disorder shape the localisation
length and the invariant measure, it is useful to apply the general
expressions~(\ref{genrho}) and~(\ref{lyap}) to specific forms of correlated
disorder. We devote the rest of this paper to this task.

\section{The case of uncorrelated disorder}
\label{uncorrelated_disorder}

\begin{figure}[hbt]
\begin{center}
\includegraphics[width=5in,height=3.5in]{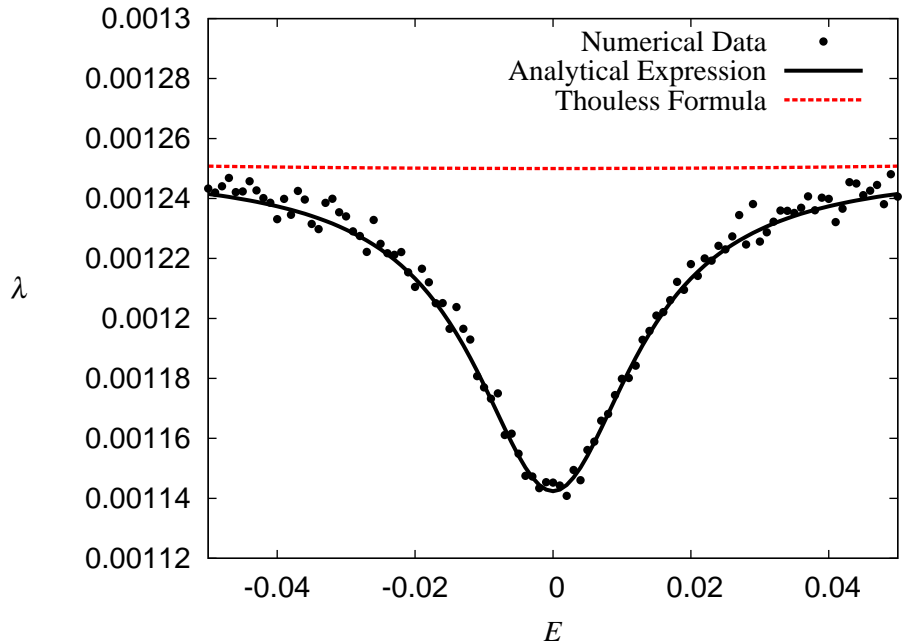}
\caption{\label{lyap_uncorr}
Inverse localisation length $\lambda$ versus $E$ in a neighbourhood of the
band centre for uncorrelated disorder. The points correspond to the numerical
data; the solid line to Eq.~(\ref{lyap_uncor}); the dashed line to Thouless'
formula. The data were obtained for $\sigma^{2} = 10^{-2}$.}
\end{center}
\end{figure}

As a first application of the general formulae obtained in
Sec.~\ref{general_formulae}, we consider the limit case of uncorrelated
disorder.
When the site energies $\varepsilon_{n}$ are {\em independent} random variables,
the binary correlator~(\ref{bincor}) reduces to a Kronecker delta
\begin{displaymath}
\chi_{0}(l) = \delta_{l0} 
\end{displaymath}
and the power spectrum~(\ref{power_spectrum}) takes a constant
unitary value.
The invariant distribution $\rho$ is obtained from the general
expression~(\ref{genrho}) with $W(0) = W(\pi/2) = 1$. One has
\begin{equation}
\begin{array}{ccl}
\rho(\theta,\varkappa) & = & \displaystyle
\frac{1}{N(\varkappa)}
\frac{e^{-8 \varkappa F_{0}(\theta)}}{\sqrt{3 + \cos(4\theta)}}
\left[e^{4 \sqrt{2} \pi \varkappa}
\int_{0}^{\theta} \frac{e^{8 \varkappa F_{0}(\phi)}}{\sqrt{3 + \cos(4\phi)}}
\mathrm{d}\phi \right. \\
& + & \displaystyle \left.
\int_{\theta}^{2 \pi} \frac{e^{8 \varkappa F_{0}(\phi)}}{\sqrt{3 + \cos(4\phi)}}
\mathrm{d}\phi \right] ,
\end{array}
\label{rho_uncor}
\end{equation}
where $N(\varkappa)$ is the normalisation factor and the function
$F_{0}(\theta)$ is defined by the integral expression
\begin{equation}
F_{0}(\theta) = \int_{0}^{\theta} \mathrm{d} \phi 
\frac{1}{3 + \cos (4 \phi)} .
\label{f0}
\end{equation}
Eq.~(\ref{rho_uncor}) coincides with the result originally obtained
in~\cite{Tes12}.
Taking into account that $W(\pi/2) = 1$ for uncorrelated disorder, the
general expression~(\ref{lyap}) for the inverse localisation length
reduces to
\begin{equation}
\lambda (\varkappa) = \frac{\sigma^{2}}{8} \left[ 1 + \langle \cos (4 \theta)
\rangle \right] ,
\label{lyap_uncor}
\end{equation}
where the average of the cosine function must be computed using the
distribution~(\ref{rho_uncor}).
In Fig.~\ref{lyap_uncorr} we compare the numerically computed Lyapunov
exponent with the expression~(\ref{lyap_uncor}) and with Thouless' formula.

At the band centre, the invariant distribution~(\ref{bcrho}) and the
inverse localisation length~(\ref{bclyap}) assume the well-known
forms~\cite{Izr98}
\begin{equation}
\rho(\theta,\varkappa=0) = \frac{1}{\mathbf{K}(1/\sqrt{2})}
\frac{1}{\sqrt{3 + \cos (4 \theta)}} ,
\label{bcrho_uncor}
\end{equation}
and
\begin{equation}
\lambda(\varkappa = 0) = \frac{\sigma^{2}}{4} \left[ 2
\frac{\mathbf{E}\left( 1/\sqrt{2} \right)}
{\mathbf{K}\left( 1/\sqrt{2} \right)} - 1 \right] =
\sigma^{2} \left[ \frac{\Gamma \left( 3/4 \right)}{\Gamma \left( 1/4 \right)}
\right]^{2} .
\label{bclyap_uncor}
\end{equation}
In conclusion, the general expressions~(\ref{genrho}) and~(\ref{lyap})
reproduce the correct results in the limit case of uncorrelated disorder.

\section{Disorder with exponentially decaying, positive correlations}
\label{exp_dec}

In this section and in those that follow we consider the Anderson
model~(\ref{andmod}) with correlated site energies. Sequences of
random variables $\{\varepsilon_{n}\}$ with arbitrary binary correlations
can be generated with the usual technique of filtering sequences
of uncorrelated random variables~\cite{Izr12}.

We now focus our attention on the case of random site energies with
positive spatial correlations that decay exponentially with the distance
between sites.
The key result is that the band-centre anomaly is quickly suppressed
for increasing values of the correlation length, as first observed
in~\cite{Her15}. As for the localisation length, it increases linearly
with the correlation length, $l_{\mathrm{loc}} \propto l_{c}$.

\begin{figure}[htb]
\begin{center}
\includegraphics[width=5in,height=3.5in]{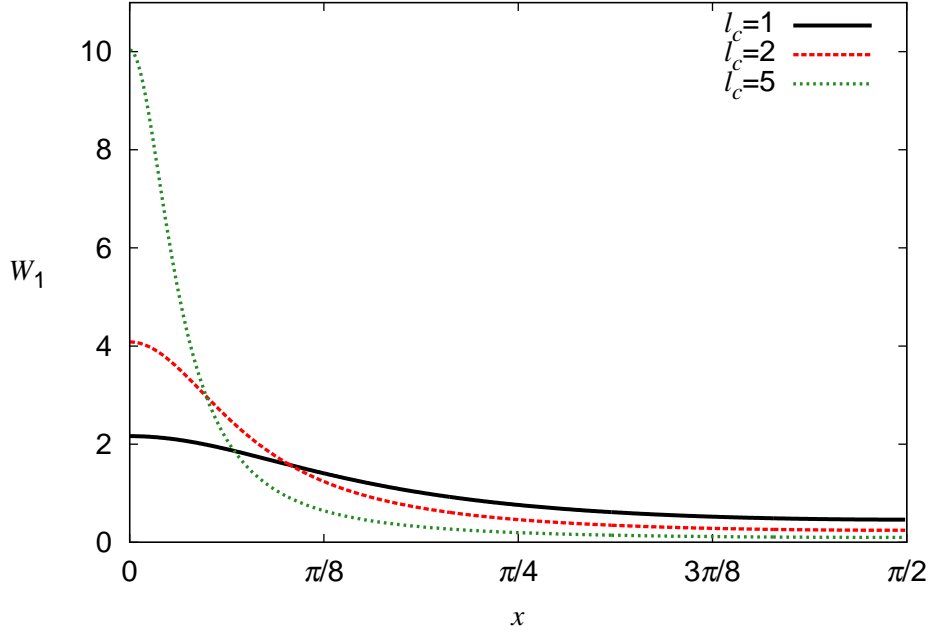}
\caption{\label{ps_ed}
Power spectrum~(\ref{ps_expdec}) versus $x$ for various values of the
correlations length.}
\end{center}
\end{figure}

We consider a binary correlator~(\ref{bincor}) of the form
\begin{equation}
\chi_{1}(l) = e^{-|l|/l_{c}} .
\label{expdec}
\end{equation}
The corresponding power spectrum~(\ref{power_spectrum}) is
\begin{equation}
W_{1}(x) = \sum_{l=-\infty}^{\infty} \chi(l) e^{i2lx} =
\frac{\sinh(1/l_{c})}{\cosh(1/l_{c}) - \cos(2x)} .
\label{ps_expdec}
\end{equation}
The behaviour of the power spectrum~(\ref{ps_expdec}) is represented in
Fig.~\ref{ps_ed} for various values of the correlation length.
The values of the power spectrum~(\ref{ps_expdec}) at the boundaries of its
domain are
\begin{equation}
W_{1}(0) = \frac{1}{W_{1}(\pi/2)} = \frac{\sinh(1/l_{c})}{\cosh(1/l_{c}) - 1} .
\label{ps_boundary_values_ed}
\end{equation}
Note that $W_{1}(0)$ and $W_{1}(\pi/2)$ are, respectively, an increasing and a
decreasing function of $l_{c}$.

\begin{figure}[htb]
\begin{center}
\includegraphics[width=5in,height=3.5in]{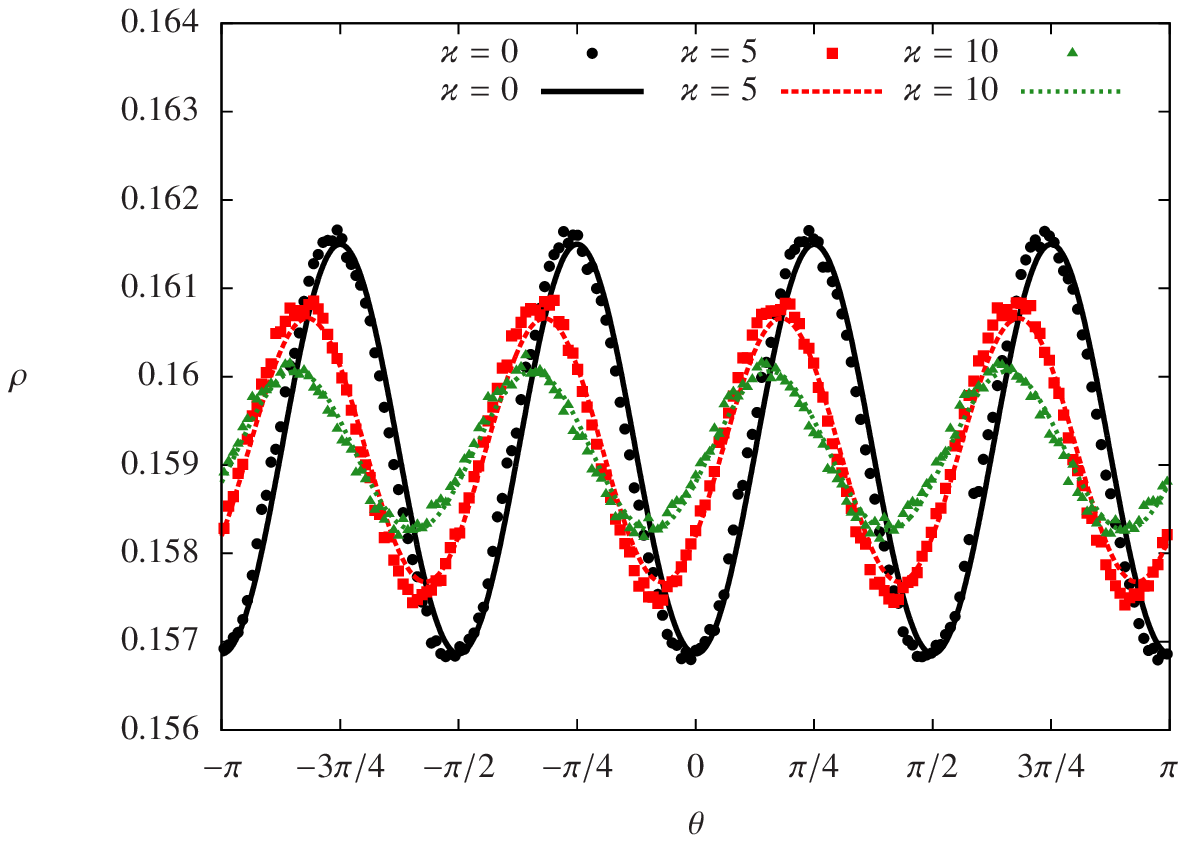}
\caption{\label{invdis_ed_kappa}
Invariant distribution $\rho(\theta,\varkappa)$ for various values of
$\varkappa$. The disorder exhibits positive exponential
correlations of the form~(\ref{expdec}) with $l_{c} = 2$. Lines
correspond to the theoretical formula~(\ref{genrho}); points to the
numerical data.
The numerical data were obtained for $\sigma^{2} = 10^{-3}$.}
\end{center}
\end{figure}

\begin{figure}[thb]
\begin{center}
\includegraphics[width=5in,height=3.5in]{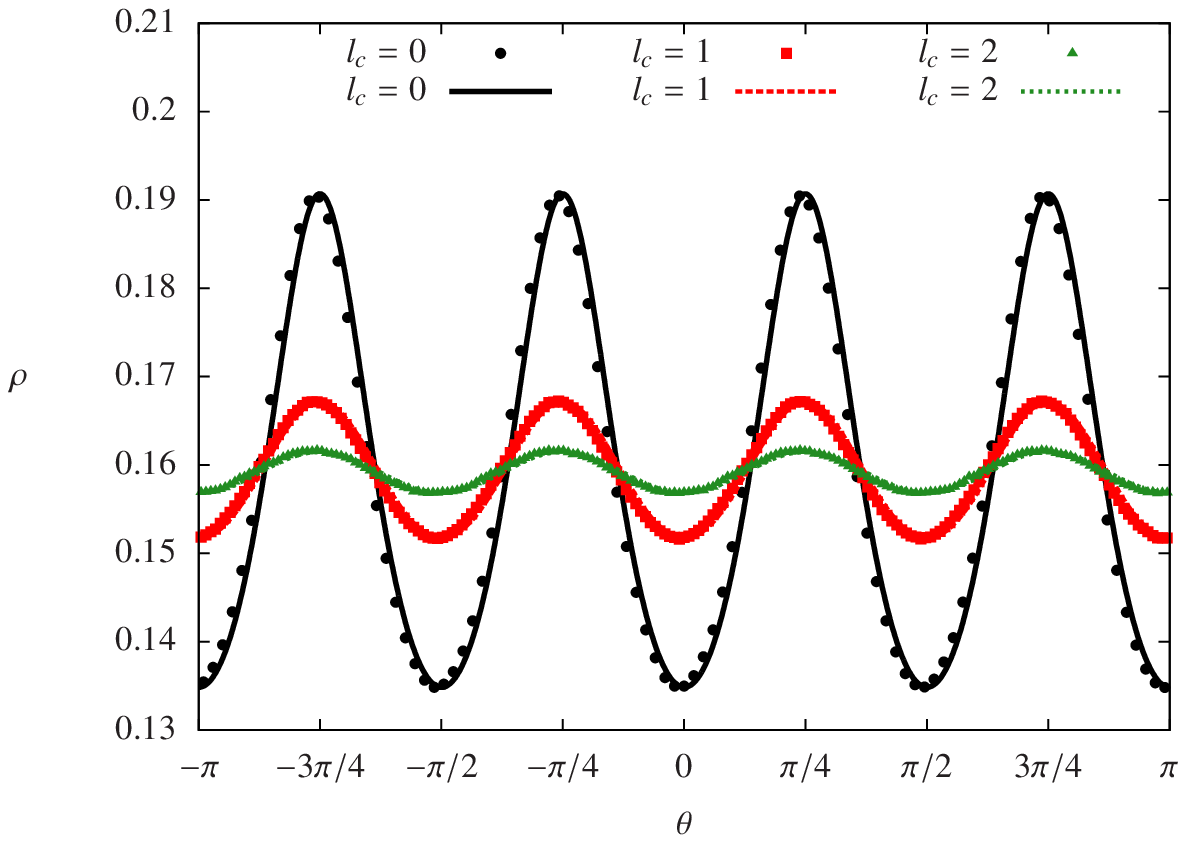}
\caption{\label{invdis_ed_lc}
Invariant distribution $\rho(\theta,\varkappa=0)$ at the band centre for
various values of the correlation length $l_{c}$. The disorder exhibits
positive exponential correlations of the form~(\ref{expdec}). The lines
correspond to the theoretical formula~(\ref{bcrho_ed}); the points to
the numerical data. The data were obtained for $\sigma^{2} = 10^{-3}$.}
\end{center}
\end{figure}

The invariant distribution for the angle variable is obtained by
substituting the values~(\ref{ps_boundary_values_ed}) in the general
expression~(\ref{genrho}). The analytical prediction matches well the
numerical results, as can be seen in Figs.~\ref{invdis_ed_kappa}
and~\ref{invdis_ed_lc}.
Fig.~\ref{invdis_ed_kappa} shows how the resonance effect is
progressively reduced as the energy moves away from the band centre.
This effect is expected and is found also in the case of uncorrelated
disorder~\cite{Tes12}.

\begin{figure}[htb]
\begin{center}
\includegraphics[width=5in,height=3.5in]{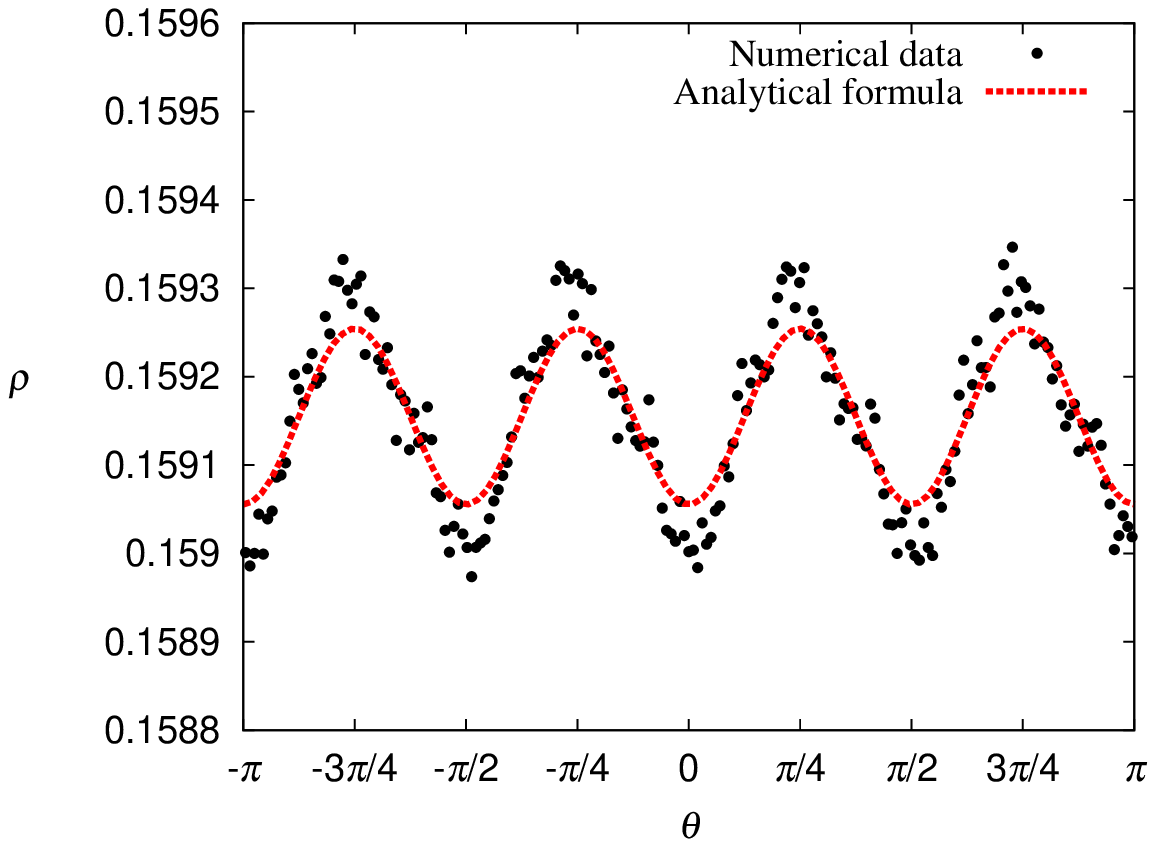}
\caption{\label{invdis_ed_large_lc}
Invariant distribution $\rho(\theta,\varkappa=0)$ at the band centre for
disorder with correlations of the form~(\ref{expdec}) with $l_{c} = 10$.
The line corresponds to the asymptotic formula~(\ref{bcrho_ed_large_lc});
the points to the numerical data. The data were obtained for
$\sigma^{2} = 10^{-3}$.}
\end{center}
\end{figure}

More surprising is the decrease of the modulation of the invariant
measure~(\ref{genrho}) with the correlation length. The suppression of
the band-centre anomaly as $l_{c}$ is increased is particularly evident
at the band-centre, where the invariant distribution~(\ref{bcrho})
takes the form
\begin{equation}
\rho(\theta,\varkappa = 0) = 
\frac{1}{\displaystyle \mathbf{K} \left( \sqrt{\varphi_{-}(l_{c})/2} \right)}
\frac{1}{\displaystyle
\sqrt{ 4  - \varphi_{-}(l_{c})
\left[ 1 - \cos \left( 4 \theta \right) \right] }}
\label{bcrho_ed}
\end{equation}
with
\begin{displaymath}
\varphi_{-}(l_{c}) =  1 - \frac{1}{\cosh \left( 1/l_{c} \right)} .
\end{displaymath}
For $l_{c} \to 0$, expression~(\ref{bcrho_ed}) differs from the corresponding
formula~(\ref{bcrho_uncor}) for uncorrelated disorder by exponentially small
terms of order $O(e^{-1/l_{c}})$.
As $l_{c}$ increases, however, the modulation of $\rho$ diminishes quickly,
as shown by Fig.~\ref{invdis_ed_lc}.
To understand this effect, it is useful to observe that, for $l_{c} \gg 1$,
expression~(\ref{bcrho_ed}) reduces to
\begin{equation}
\rho(\theta, \varkappa = 0) = \frac{1}{2 \pi} \left[ 1 - \frac{1}{16 l_{c}^{2}}
\cos \left( 4 \theta \right) + \ldots \right] .
\label{bcrho_ed_large_lc}
\end{equation}
The asymptotic form~(\ref{bcrho_ed_large_lc}) matches relatively well
the numerical data already for values of $l_{c} \gtrsim 10$, as shown by
Fig.~\ref{invdis_ed_large_lc}.

\begin{figure}[hbt]
\begin{center}
\includegraphics[width=5in,height=3.5in]{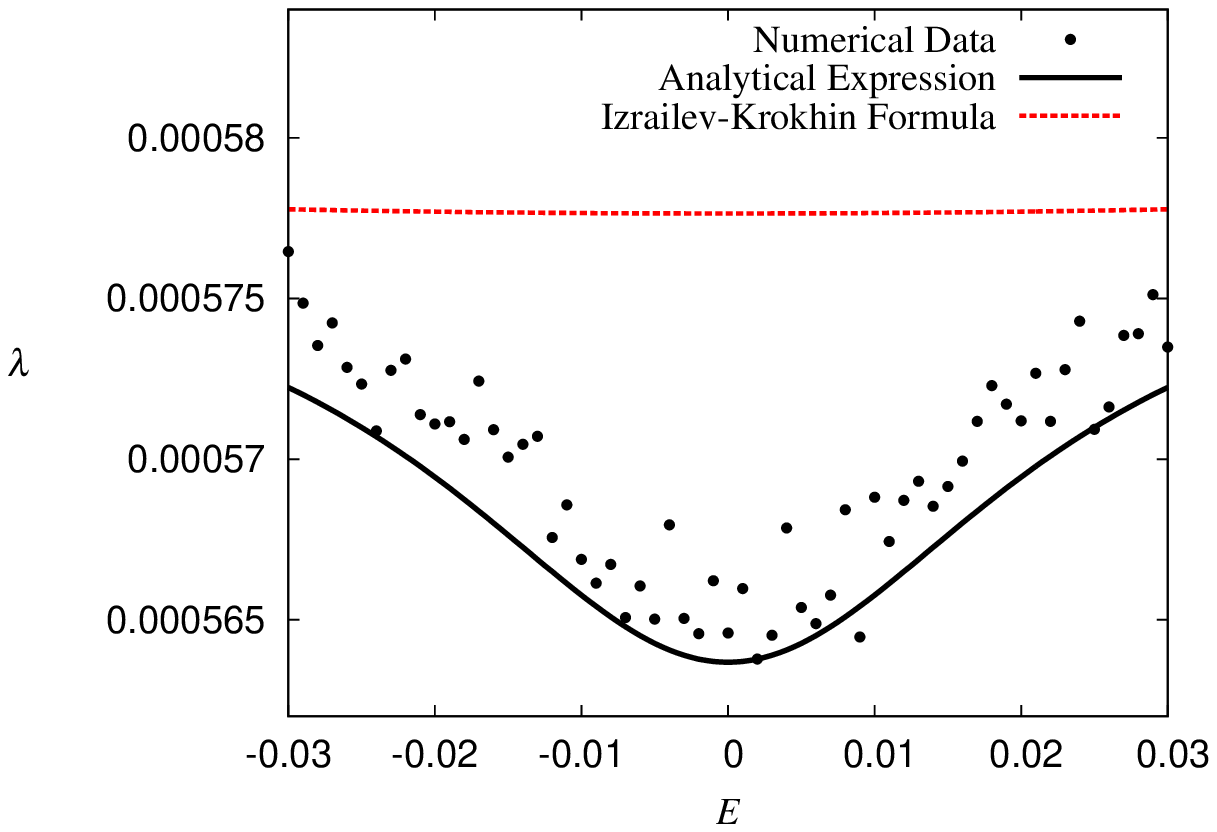}
\caption{\label{lyap_ed_lc_1}
Inverse localisation length $\lambda$ versus $E$ in a neighbourhood of the
band centre for disorder with correlations of the form~(\ref{expdec})
and $l_{c} = 1$.
The points correspond to the numerical data; the solid line to
Eq.~(\ref{lyap}); the dashed line to the formula~(\ref{nonres_lyap}).
The numerical data were obtained for
$\sigma^{2} = 10^{-2}$.}
\end{center}
\end{figure}

The inverse localisation length can be computed with the help of
Eqs.~(\ref{genrho}) and~(\ref{lyap}), supplemented by the specific
values~(\ref{ps_boundary_values_ed}) of the power spectrum.
In Fig.~\ref{lyap_ed_lc_1} we compare the numerically computed Lyapunov
exponent with the theoretical expression~(\ref{lyap}) and with the
formula~(\ref{nonres_lyap}) obtained by Izrailev and Krokhin.
Note that the differences between the numerical values of $\lambda$ and
the values predicted by Eq.~(\ref{lyap}) are of order $\sim 10^{-5}$, well
within the $O(\sigma^{4})$ error intrinsic to the second-order approximation
used in the theoretical calculations.

At the exact band centre, a relatively simple analytical expression
for the inverse localisation length can be obtained by substituting
the values~(\ref{ps_boundary_values_ed}) in Eq.~(\ref{bclyap}).
One obtains
\begin{equation}
\lambda (\varkappa = 0) = \frac{\sigma^{2}}{4 \sinh \left( 1/l_{c} \right)}
\left\{ 2 \cosh \left( 1/l_{c} \right)
\frac{\displaystyle \mathbf{E} \left(\sqrt{\varphi_{-}(l_{c})/2} \right)}
{\displaystyle \mathbf{K} \left(\sqrt{\varphi_{-}(l_{c})/2} \right)} -
\cosh \left( 1/l_{c} \right) - 1 \right\} .
\label{bclyap_ed}
\end{equation}
Note that, for $l_{c} \gg 1$, the inverse localisation
length~(\ref{bclyap_ed}) becomes
\begin{displaymath}
\lambda (\varkappa = 0) \simeq \frac{\sigma^{2}}{16 l_{c}} .
\end{displaymath}
Therefore the extension of the band-centre state increases linearly with
the correlation length $l_{c}$.

\section{Exponentially decaying correlations with oscillating sign}
\label{oscexpdec}

We now consider correlations which decay exponentially with the distance
between sites, but whose sign oscillates. Contrary to the previous case,
the anomaly is now reinforced as the correlation length $l_{c}$ increases.

\begin{figure}[htb]
\begin{center}
\includegraphics[width=5in,height=3.5in]{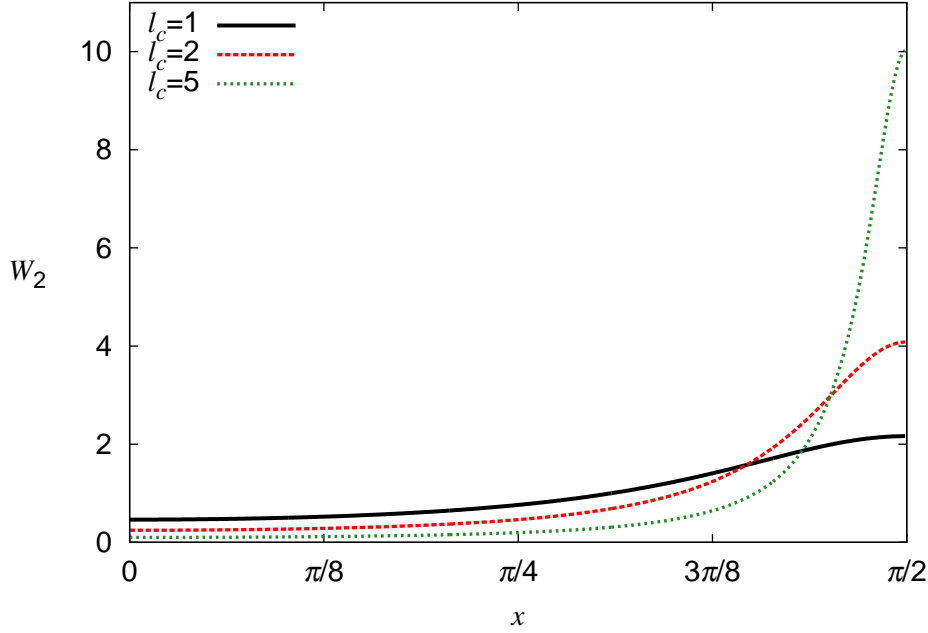}
\caption{\label{ps_as}
Power spectrum~(\ref{ps_expdec_as}) versus $x$ for various values of the
correlations length.}
\end{center}
\end{figure}

Mathematically, the correlations of the site energies have the form
\begin{equation}
\chi_{2}(l) = (-1)^{l} e^{-|l|/l_{c}} .
\label{expdec_as}
\end{equation}
The corresponding power spectrum is
\begin{equation}
W_{2}(x) = \sum_{l=-\infty}^{\infty} \chi(l) e^{i2lx} =
\frac{\sinh(1/l_{c})}{\cosh(1/l_{c}) + \cos(2x)} .
\label{ps_expdec_as}
\end{equation}
The behaviour of the power spectrum~(\ref{ps_expdec_as}) is represented
in Fig.~\ref{ps_as}.
Eq.~(\ref{ps_expdec_as}) implies that
\begin{equation}
W_{2}(0) = \frac{1}{W_{2}(\pi/2)} = \frac{\sinh(1/l_{c})}{\cosh(1/l_{c}) + 1} .
\label{ps_boundary_values_as}
\end{equation}
Note that
\begin{displaymath}
W_{2}\left( x \right) = W_{1} \left( \frac{\pi}{2} - x \right)
\end{displaymath}
and that, therefore, the formulae for the invariant distribution and the
inverse localisation length for this case can be obtained from the expressions
for the previous case with the exchanges $W_{1}(0) \leftrightarrow W_{2}(\pi/2)$
and $W_{1}(\pi/2) \leftrightarrow W_{2}(0)$.

\begin{figure}[htb]
\begin{center}
\includegraphics[width=5in,height=3.5in]{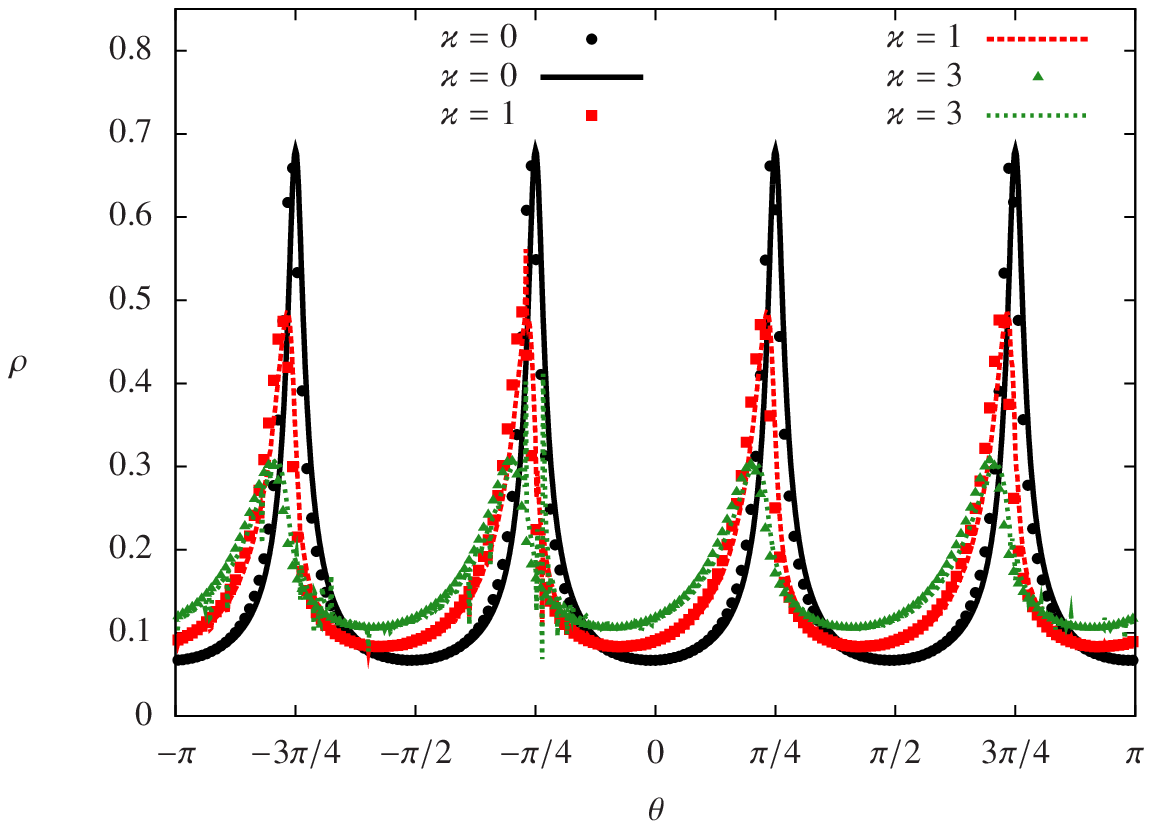}
\caption{\label{invdis_as_kappa}
Invariant distribution $\rho(\theta,\varkappa)$ for various values of the
parameter $\varkappa$. The lines correspond to the theoretical
formula~(\ref{genrho}); the points to the numerical data. The data
were obtained for disorder with correlation of the form~(\ref{expdec_as})
with $l_{c} = 5$.
The disorder strength in numerical calculations was $\sigma^{2} = 10^{-3}$.}
\end{center}
\end{figure}

\begin{figure}[htb]
\begin{center}
\includegraphics[width=5in,height=3.5in]{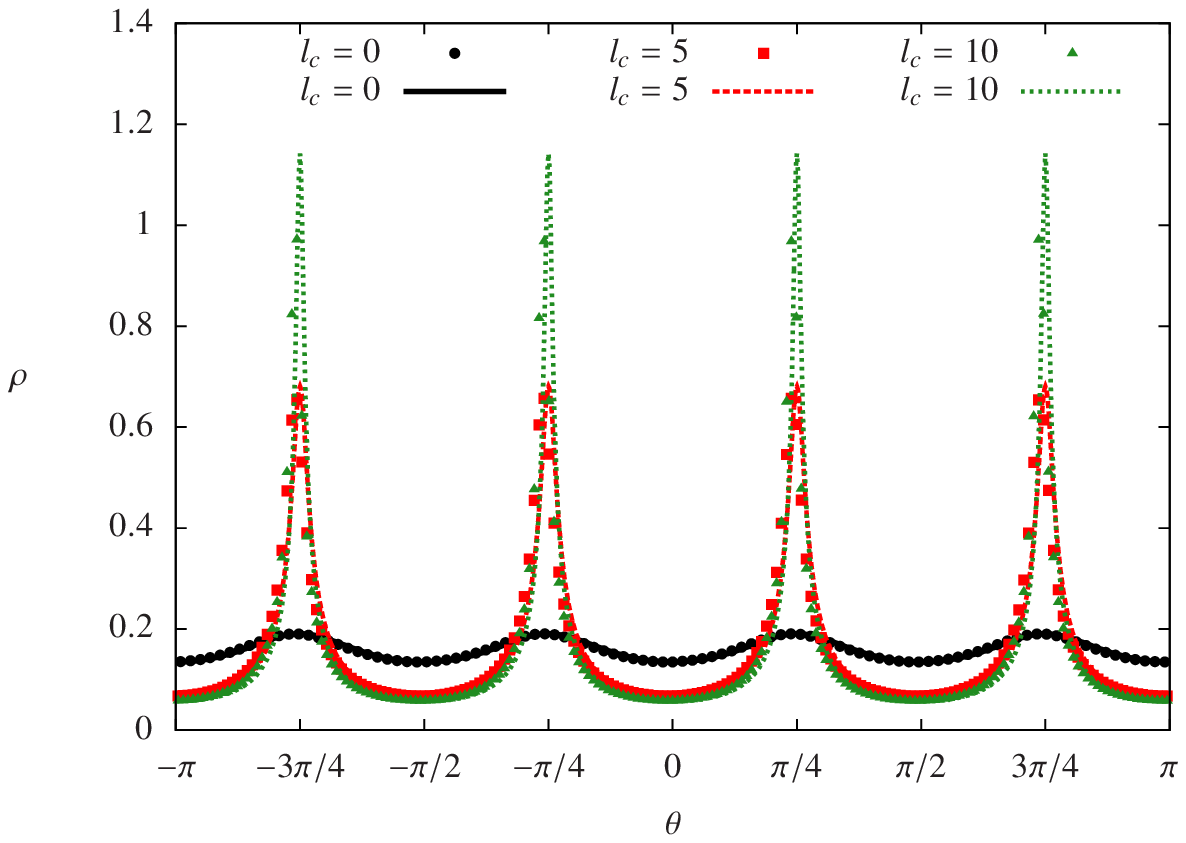}
\caption{\label{invdis_as_lc}
Invariant distribution $\rho(\theta,\varkappa=0)$ at the band centre
for various values of the correlation length $l_{c}$. The disorder exhibits
correlations of the form~(\ref{expdec_as}). The lines correspond
to the theoretical formula~(\ref{bcrho_as}); the points to the numerical data.
The data were obtained for $\sigma^{2} = 10^{-3}$.}
\end{center}
\end{figure}

As before, the invariant distribution for the angle variable is obtained by
substituting the specific values~(\ref{ps_boundary_values_as}) of $W(0)$ and
$W(\pi/2)$ in the general expression~(\ref{genrho}).
The resulting analytical formula is corroborated by the numerical results, as
shown by Figs.~\ref{invdis_as_kappa} and~\ref{invdis_as_lc}.
Fig.~\ref{invdis_as_kappa} shows how correlation of the form~(\ref{expdec_as})
produce a strong modulation of the invariant measure at the band centre
which is gradually reduced as the energy moves away from the band
centre.

Fig.~\ref{invdis_as_lc}, on the other hand, shows how the invariant
distribution develops conspicuous peaks as $l_{c}$ is increased.
This behaviour is entirely consistent with the theoretical predictions.
At the band centre, in fact, the general expression~(\ref{bcrho}) for the
invariant distribution takes the form
\begin{equation}
\rho(\theta, \varkappa = 0) =
\frac{1}{\displaystyle \mathbf{K} \left( \sqrt{ \varphi_{+}(l_{c})/2} \right)}
\frac{1}{\displaystyle
\sqrt{ 4 - \varphi_{+}(l_{c})
\left[ 1 - \cos \left( 4 \theta \right) \right] }}
\label{bcrho_as}
\end{equation}
with
\begin{displaymath}
\varphi_{+}(l_{c}) = 1 + \frac{1}{\cosh(1/l_{c})} .
\end{displaymath}
The asymptotic behaviour of the distribution~(\ref{bcrho_as}) for
$l_{c} \gg 1$ is
\begin{equation}
\rho(\theta, \varkappa = 0) \simeq \frac{1}{\displaystyle
2^{3/2} \mathbf{K} \left( \sqrt{1 - \frac{1}{4l_{c}^{2}}} \right)}
\frac{1}{\displaystyle \sqrt{1 + \frac{1}{4l_{c}^{2}} + \left( 1 -
\frac{1}{4l_{c}^{2}} \right) \cos \left( 4 \theta \right)}} .
\label{bcrho_as_large_lc}
\end{equation}
Note that for
\begin{displaymath}
\begin{array}{ccc}
\displaystyle
\theta \simeq \frac{\pi}{4} + n \frac{\pi}{2} & \mbox{ with } &
n \in \mathbf{Z}
\end{array}
\end{displaymath}
the distribution~(\ref{bcrho_as_large_lc}) assumes the value
\begin{displaymath}
\rho \left( \frac{\pi}{4}, \varkappa = 0 \right) \simeq
\frac{l_{c}}{\displaystyle
2 \mathbf{K} \left( \sqrt{1 - \frac{1}{l_{c}^{2}}} \right)}
\end{displaymath}
which diverges for $l_{c} \gg 1$. This entails that the
distribution~(\ref{bcrho_as}) develops four sharp maxima as $l_{c}$ is
increased, in agreement with the data of Fig.~\ref{invdis_as_lc}.

\begin{figure}[hbt]
\begin{center}
\includegraphics[width=5in,height=3.5in]{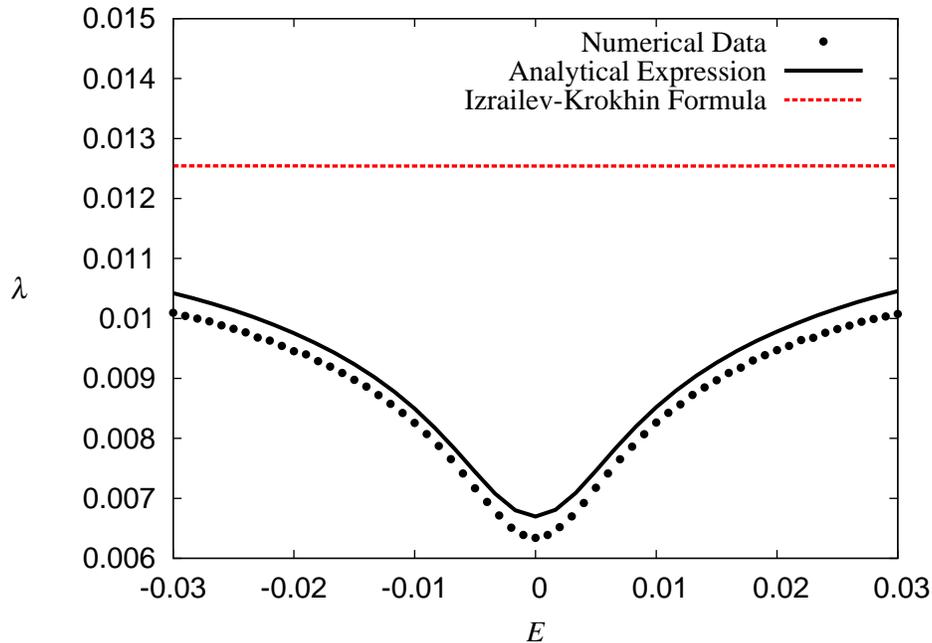}
\caption{\label{lyap_as_lc_5}
Inverse localisation length $\lambda$ versus $E$ in a neighbourhood of the
band centre for disorder with correlations of the form~(\ref{expdec_as})
with $l_{c} = 5$.
The points correspond to the numerical data; the solid line to
Eq.~(\ref{lyap}); the dashed line to the formula~(\ref{nonres_lyap}).
The numerical data were obtained for
$\sigma^{2} = 10^{-2}$.}
\end{center}
\end{figure}

After inserting the values~(\ref{ps_boundary_values_as}) in the
expression~(\ref{genrho}) for the invariant distribution, one can
evaluate the rhs of Eq.~(\ref{lyap}) and obtain the inverse localisation
length. The results agree with the numerical data, as shown by
Fig.~\ref{lyap_as_lc_5}.
The slight discrepancy between numerical data and the theoretical
expression is of order $O(\sigma^{4})$ and should probably be attributed to
the neglected fourth-order correction in Eq.~(\ref{lyap}).
Fig.~\ref{lyap_as_lc_5} graphically shows that correlations of the
form~(\ref{expdec_as}) increase the {\em relative} deviation of the
Lyapunov exponent from the value predicted by the formula~(\ref{nonres_lyap})
obtained by Izrailev and Krokhin. This is consistent with the very large
peaks that the invariant distribution develops in the present case.

Correlations of the form~(\ref{expdec_as}) have also the effect of
enhancing the localisation of the electronic states for increasing
values of the correlation length.
This is a consequence of the fact that the factor $W(\pi/2)$, defined
by Eq.~(\ref{ps_boundary_values_as}), is an increasing function of $l_{c}$.
This conclusion is confirmed by the explicit formula for the inverse
localisation length at the exact band centre.
For $E = 0$, Eq.~(\ref{bclyap}) becomes
\begin{equation}
\lambda (\varkappa = 0) = \frac{\sigma^{2}}{4 \sinh \left( 1/l_{c} \right)}
\left\{ 2 \cosh \left( 1/l_{c} \right)
\frac{\displaystyle \mathbf{E} \left(\sqrt{ \varphi_{+}(l_{c})/2} \right)}
{\displaystyle \mathbf{K} \left(\sqrt{ \varphi_{+}(l_{c})/2 } \right)} -
\cosh \left( 1/l_{c} \right) + 1 \right\} .
\label{bclyap_as}
\end{equation}
As in the previous case, in the limit $l_{c} \to 0$ Eq.~(\ref{bclyap_as})
differs from its counterpart~(\ref{bclyap_uncor}) only by vanishing terms
of order $O(e^{-1/l_{c}})$.
In the limit $l_{c} \gg 1$, on the other hand, the inverse localisation
length~(\ref{bclyap_as}) diverges as
\begin{displaymath}
\lambda (\varkappa = 0) = \frac{\sigma^{2}}{4} \left[ \frac{2l_{c}}{\ln (8l_{c})}
- \frac{1}{2l_{c}} + \ldots \right] .
\end{displaymath}
In this case the band-centre state becomes strongly localised as $l_{c}$
increases.

\section{The case of a composite lattice}
\label{longrangedcor}

We now consider the case of a 1D chain with random energies whose binary
correlator satisfies the condition
\begin{equation}
\begin{array}{ccc}
\chi(2l+1) = 0 & \mbox{ for } & l \in \mathbf{Z} .
\end{array}
\label{twolat}
\end{equation}
A sequence $\{\varepsilon_{n}\}$ with correlations of the form~(\ref{twolat})
can be obtained by mixing two independent random sequences $\{\alpha_{n}\}$
and $\{\beta_{n}\}$ with the same statistical properties.
More precisely, one assumes that
\begin{displaymath}
\langle \alpha_{n} \rangle = \langle \beta_{n} \rangle = 0
\end{displaymath}
and that the binary correlators are
\begin{displaymath}
\begin{array}{ccc}
\langle \alpha_{n} \alpha_{n+l} \rangle =
\langle \beta_{n} \beta_{n+l} \rangle =
\sigma^{2} \tilde{\chi}(l) &
\mbox{ and } &
\langle \alpha_{n} \beta_{m} \rangle = 0 ,
\end{array}
\end{displaymath}
where $\tilde{\chi}(l)$ is an arbitrary function, satisfying the condition
that is quickly decays for $l \gg l_{c}$.
One can then define the random energies by setting
\begin{displaymath}
\begin{array}{ccc}
\varepsilon_{2n} = \alpha_{n} & \mbox{ and } &
\varepsilon_{2n+1} = \beta_{n} .
\end{array}
\end{displaymath}
In physical terms, the chain is split in two independent and interpenetrating
sublattices.

Condition~(\ref{twolat}) implies that the power spectrum must satisfy the
identity
\begin{displaymath}
W(0) = W \left( \frac{\pi}{2} \right) .
\end{displaymath}
As a consequence, the invariant distribution~(\ref{genrho}) assumes the
particularly simple form
\begin{equation}
\begin{array}{ccl}
\rho(\theta,\varkappa) & = & \displaystyle
\frac{1}{N(\tilde{\varkappa})}
\frac{e^{-8 \tilde{\varkappa} F_{0}(\theta)}}{\sqrt{3 + \cos(4\theta)}}
\left[e^{4 \sqrt{2} \pi \tilde{\varkappa}}
\int_{0}^{\theta} \frac{e^{8 \tilde{\varkappa} F_{0}(\phi)}}{\sqrt{3 + \cos(4\phi)}}
\mathrm{d}\phi \right. \\
& + & \displaystyle \left.
\int_{\theta}^{2 \pi} \frac{e^{8 \tilde{\varkappa} F_{0}(\phi)}}{\sqrt{3 + \cos(4\phi)}}
\mathrm{d}\phi \right] ,
\end{array}
\label{rho_twolat}
\end{equation}
where $N$ is a normalisation constant, $F_{0}(\theta)$ is
defined by Eq.~(\ref{f0}), and
\begin{equation}
\tilde{\varkappa} = \frac{\varkappa}{W(0)} .
\label{res_kappa}
\end{equation}
The distribution~(\ref{rho_twolat}) has the same form of the
distribution~(\ref{rho_uncor}) for uncorrelated disorder, the only
difference being that the parameter~(\ref{kappa}) of the latter is replaced
by the rescaled parameter~(\ref{res_kappa}) in the former.
In physical terms, this means that the correlations do not modify the
modulation of $\rho$, but they alter the scale over which the
distance of the energy from the band centre is measured.
This entails that the anomaly extends over a larger energy interval if
$W(0) > 1$ and is restricted to a shrunken region if $W(0) < 1$.

At the exact band centre, $\tilde{\varkappa} = \varkappa = 0$ and the
invariant distribution reduces to the form~(\ref{bcrho_uncor}).
Correspondingly, the inverse localisation length becomes
\begin{equation}
\lambda(\varkappa = 0) = \frac{\sigma^{2}}{4} \left[ 2
\frac{\mathbf{E}\left( 1/\sqrt{2} \right)}
{\mathbf{K}\left( 1/\sqrt{2} \right)} - 1 \right]
W\left( \frac{\pi}{2} \right) .
\label{bclyap_twolat}
\end{equation}

\subsection{Long-ranged correlations}

\begin{figure}[htb]
\begin{center}
\includegraphics[width=5in,height=3.5in]{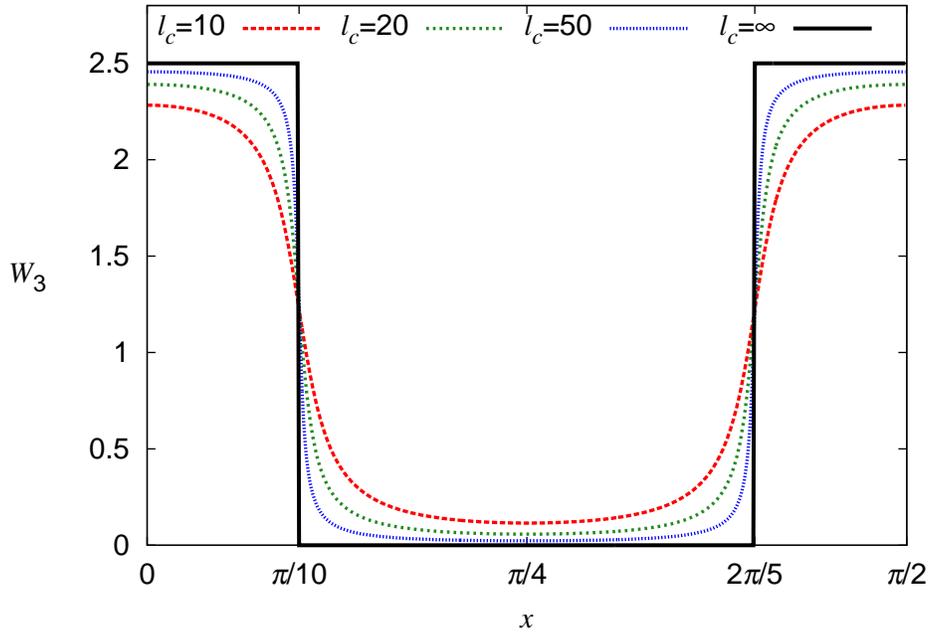}
\caption{\label{ps_longranged}
Power spectrum~(\ref{ps_lr}) with $a = \pi/10$ for various values of the
correlations length. For $l_{c} \to \infty$, mobility edges appear
at $x_{1} = \pi/10$ and $x_{2} = 2 \pi/5$.}
\end{center}
\end{figure}

We now focus our attention on a specific type of correlator fulfilling
condition~(\ref{twolat}), i.e.,
\begin{equation}
\chi_{3}(l) = \frac{1+(-1)^{l}}{2} \frac{\sin \left( 2al \right)}{2al} ,
\label{chi_lr}
\end{equation}
where the parameter $a$ lies in the interval $[0,\pi/4]$.
The correlator~(\ref{chi_lr}) does not decreases quickly for $l \gtrsim l_{c}$,
and therefore does not satisfy one of the conditions used to derive our
analytical results. It can be fitted in our theoretical framework, however,
if it considered as the limit form for $l_{c} \to \infty$ of the correlator
\begin{equation}
\chi_{3}(l) = \frac{1+(-1)^{l}}{2} \frac{\sin \left( 2al \right)}{2al}
\exp \left(-\frac{|l|}{l_{c}} \right) .
\label{bincor_lr}
\end{equation}
The power spectrum of this correlator is
\begin{equation}
\begin{array}{ccl}
W_{3}(x) & = & \displaystyle 1 + \frac{1}{4a} \left\{
\arctan \left[ \frac{\sin \left( 2a + 2x \right)}
{e^{1/l_{c}} - \cos \left( 2a + 2x \right)} \right] -
\arctan \left[ \frac{\sin \left( 2a + 2x \right)}
{ e^{1/l_{c}} + \cos \left( 2a + 2x \right)} \right] \right. \\
& + & \displaystyle \left.
\arctan \left[ \frac{\sin \left( 2a - 2x \right)}
{e^{1/l_{c}} - \cos \left( 2a - 2x \right)} \right] -
\arctan \left[ \frac{\sin \left( 2a - 2x \right)}
{ e^{1/l_{c}} + \cos \left( 2a - 2x \right)} \right] \right\} .\\
\end{array}
\label{ps_lr}
\end{equation}
In the limit $l_{c} \to \infty$ the power spectrum~(\ref{ps_lr}) tends to
the form
\begin{equation}
W_{3}(x) = \left\{ \begin{array}{ccl}
\displaystyle \frac{\pi}{4a} & \mbox{if} & \displaystyle
x \in \left[0, a \right] \cup \left[ \frac{\pi}{2} - a, \frac{\pi}{2} \right] \\
0 & \mbox{if} & \displaystyle
x \in \left[ a, \frac{\pi}{2} - a \right] \\
\end{array} . \right.
\label{ps_lr_infty}
\end{equation}
The behaviour of the power spectrum~(\ref{ps_lr}) with $a = \pi/10$
is represented in Fig.~\ref{ps_longranged} for various values
of $l_{c}$.
In the limit $l_{c} \to \infty$, the power spectrum vanishes
for $x_{1} = \pi/10 < x < x_{2} = 2 \pi/2$; according to the standard
formula~(\ref{nonres_lyap}), this generates mobility edges at
$E_{1} \simeq \pm 0.618$ and $E_{2} \simeq \pm 1.902$.

\begin{figure}[htb]
\begin{center}
\includegraphics[width=5in,height=3.5in]{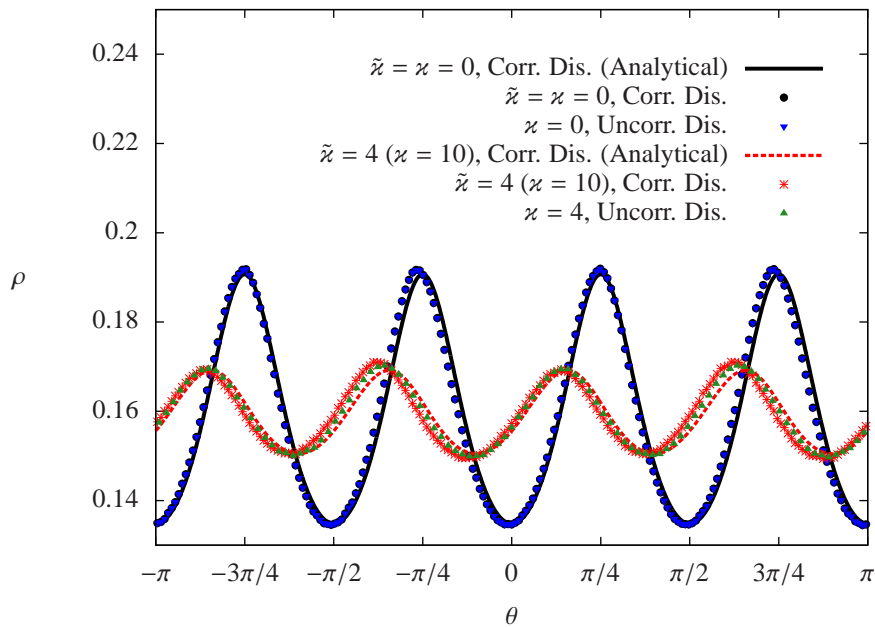}
\caption{\label{invdis_lr_kappa}
Invariant distribution $\rho(\theta,\tilde{\varkappa})$ for disorder
with correlations of the form~(\ref{chi_lr}) with $a = \pi/10$.
The data correspond to the values $\tilde{\varkappa} = 0$ (band centre)
and $\tilde{\varkappa} = 4$.
Lines correspond to the theoretical formula~(\ref{rho_twolat}); points to
the numerical data.
Also represented are the numerical data for the invariant distributions
for uncorrelated disorder and values $\varkappa = 0$ and $\varkappa = 4$
of the parameter~(\ref{kappa}).
In numerical computations the disorder strength was set at
$\sigma^{2} = 10^{-2}$.}
\end{center}
\end{figure}

When the binary correlator takes the form~(\ref{bincor_lr}), the invariant
distribution is given by the expression~(\ref{rho_twolat}).
This is true for any finite value of the correlation length; under the
reasonable assumption that $\rho(\theta,\varkappa)$ should be a continuous
function of $l_{c}$, one can conclude that the invariant distribution
keeps the form~(\ref{rho_twolat}) even in the limit $l_{c} \to \infty$,
i.e., when the binary correlator is given by Eq.~(\ref{chi_lr}).
The numerical data corroborate this conclusion, as shown by
Fig.~\ref{invdis_lr_kappa}.
As can be seen, the numerical data match well the theoretical
distribution~(\ref{rho_twolat}) both at the band centre
($\tilde{\varkappa} = 0$) and away from it ($\tilde{\varkappa} = 4$).
In Fig.~\ref{invdis_lr_kappa} we also plot the numerically obtained
invariant distributions for uncorrelated disorder with $\varkappa =0$
(band centre) and $\varkappa =4$. The data show that, as expected, these
two distributions collapse on the distributions for correlated disorder
with $\tilde{\varkappa} = 0$ and $\tilde{\varkappa} = 4$. Note that in
the present case value $W(0) = 5/2$ so that $\varkappa = 10$ when
$\tilde{\varkappa} = 4$.

\begin{figure}[hbt]
\begin{center}
\includegraphics[width=5in,height=3.5in]{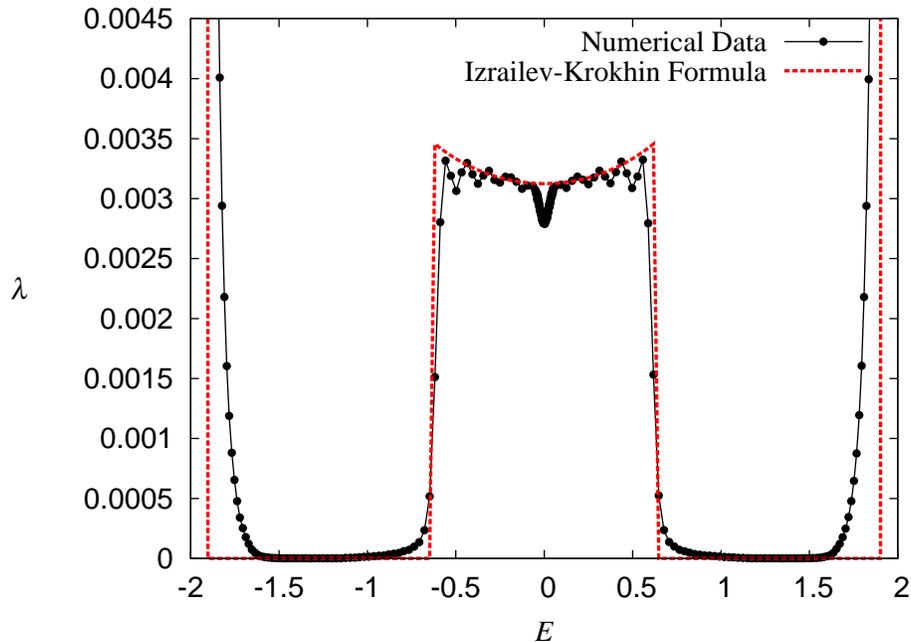}
\caption{\label{lyap_lr}
Inverse localisation length $\lambda$ versus $E$ for disorder with
correlations of the form~(\ref{chi_lr}). The points correspond
to numerical data; the dashed line to the standard
formula~(\ref{nonres_lyap}).
The data were obtained for $\sigma^{2} = 10^{-2}$.}
\end{center}
\end{figure}

\begin{figure}[hbt]
\begin{center}
\includegraphics[width=5in,height=3.5in]{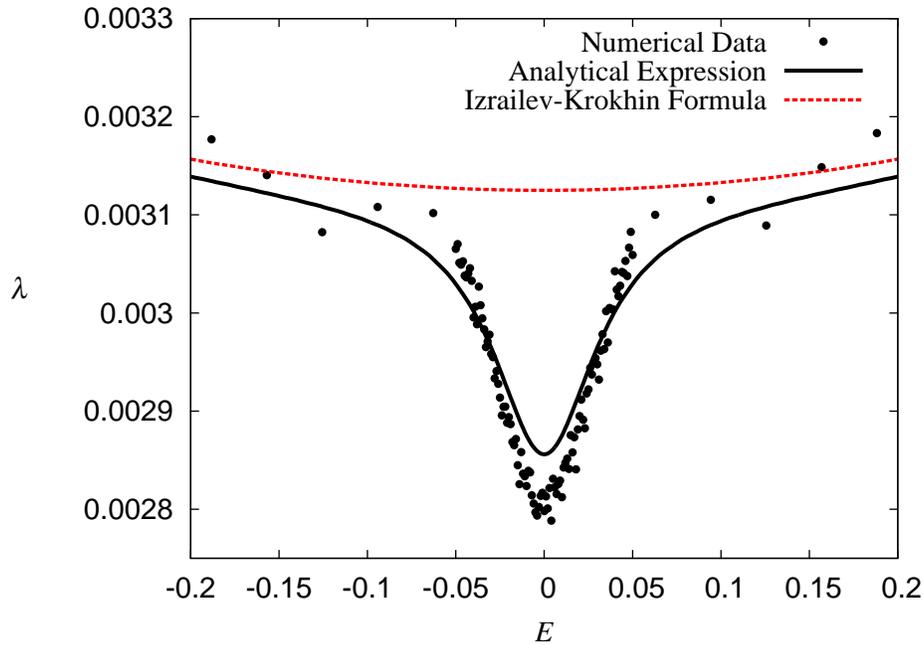}
\caption{\label{lyap_lr_nbc}
Inverse localisation length $\lambda$ versus $E$ in a neighbourhood of the
band centre for disorder with correlations of the form~(\ref{chi_lr}).
The points correspond to the numerical data; the solid line to
Eq.~(\ref{lyap3}); the dashed line to the standard
formula~(\ref{nonres_lyap}).
The data were obtained for $\sigma^{2} = 10^{-2}$.}
\end{center}
\end{figure}

We can now consider the behaviour of the localisation length.
In Fig.~\ref{lyap_lr} we plot the Lyapunov exponent as a function of the
energy. One can easily see that the formula~(\ref{nonres_lyap}) proposed
by Izrailev and Krokhin agrees well with the numerical data; in
particular, effective mobility edges arise where expected.
Discrepancies between the numerical data and the expression~(\ref{nonres_lyap})
appear at the band centre and at the band edges, however, where anomalies
occur.
The band-centre anomaly is represented in greater detail in
Fig.~\ref{lyap_lr_nbc}, where we compare the numerical data with the
standard formula~(\ref{nonres_lyap}) and with Eq.~(\ref{lyap3}).
The use of the expression~(\ref{lyap3}), rather then~(\ref{lyap}), is due
to the fact that in this case the anomaly extends, as predicted, over a
larger energy interval and, therefore, the term $\cos^{2}(\delta)$ in
Eq.~(\ref{lyap3}) cannot be approximated with unity as done so far.
Fig.~\ref{lyap_lr_nbc} confirms that our theoretical results work rather
well even for long-ranged correlations of the form~(\ref{chi_lr}).
The extension of the energy interval with anomalous behaviour can be clearly
seen if one compares the data represented in Fig.~\ref{lyap_lr_nbc} with
those corresponding to uncorrelated disorder shown in Fig.~\ref{lyap_uncorr}.

\begin{figure}[thb]
\begin{center}
\includegraphics[width=5in,height=3.5in]{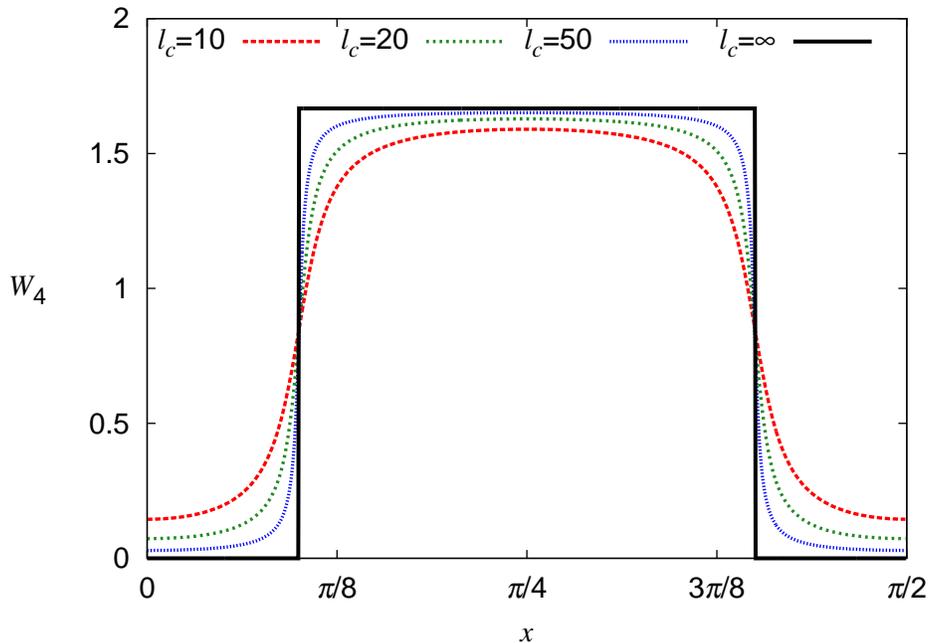}
\caption{\label{ps2_longranged}
Power spectrum~(\ref{ps2_lr}) with $a = \pi/10$ for various values of the
correlations length. As $l_{c} \to \infty$, mobility edges arise at
$x_{1} = \pi/10$ and $x_{2} = 2 \pi/5$.}
\end{center}
\end{figure}

To conclude the discussion of long-ranged correlations, we can consider
the case of disorder with correlations of the form
\begin{equation}
\chi_{4}(l) = \frac{\sin \left( 2al \right)}{\left( 2al \right)}
\cos \left( \frac{\pi}{2}l \right) \exp \left( - \frac{|l|}{l_{c}} \right). 
\label{chi2_lr}
\end{equation}
The limit form for $l_{c} \to \infty$ is
\begin{displaymath}
\chi_{4}(l) = \frac{\sin \left( 2al \right)}{\left( 2al \right)}
\cos \left( \frac{\pi}{2}l \right) .
\end{displaymath}
The power spectrum corresponding to the correlator~(\ref{chi2_lr}) is
\begin{equation}
\begin{array}{ccl}
W_{4}(x) & = & \displaystyle 1 + \frac{1}{4a} \left\{
\arctan \left[ \frac{e^{1/l_{c}} + \sin \left( 2x - 2a \right)}
{\cos \left( 2x - 2a \right)} \right] +
\arctan \left[ \frac{e^{1/l_{c}} - \sin \left( 2x + 2a \right)}
{\cos \left( 2x + 2a \right)} \right] \right.\\
&  & \displaystyle \left.
- \arctan \left[ \frac{e^{1/l_{c}} + \sin \left( 2x + 2a \right)}
{\cos \left( 2x + 2a \right)} \right] -
\arctan \left[ \frac{e^{1/l_{c}} - \sin \left( 2x - 2a \right)}
{\cos \left( 2x - 2a \right)} \right] \right\} .\\
\end{array}
\label{ps2_lr}
\end{equation}
In the limit $l_{c} \to \infty$ the power spectrum~(\ref{ps2_lr}) tends
to the form
\begin{displaymath}
W_{4}(x) = \left\{ \begin{array}{ccl}
\displaystyle \frac{\pi}{\pi - 4a} & \mbox{if} & \displaystyle
x \in \left[ \frac{\pi}{4} - a, \frac{\pi}{4} + a \right] \\
0 & \mbox{if} & \displaystyle
x \in \left[ 0, \frac{\pi}{4} - a \right] \cup \left[ \frac{\pi}{4} + a,
\frac{\pi}{2} \right] \\
\end{array} . \right.
\end{displaymath}
We thus obtain a power spectrum which is complementary with respect to the
case described by Eq.~(\ref{ps_lr_infty}).
The behaviour of the power spectrum~(\ref{ps2_lr}) with $a = \pi/10$
is represented in Fig.~\ref{ps2_longranged} for various values
of $l_{c}$.

As can be seen from Eq.~(\ref{ps2_lr}), for increasing values of $l_{c}$
the power spectrum tends to zero at the boundaries of the domain $[0,\pi/2]$.
This entails that the anomaly at the band centre becomes unstable
as $l_{c} \to \infty$. In fact, although at the exact band centre the
invariant distribution keeps the form~(\ref{bcrho_uncor}), the fact
that $W(0) = W(\pi/2) \to 0$ implies that the rescaled
parameter~(\ref{res_kappa}) grows very quickly for any infinitesimal
deviation of the energy from the band centre. Therefore the invariant
distribution becomes uniform very fast when the energy moves away from
the band centre.

We can conclude that, depending on the value of the power spectrum at the
band centre, correlations satisfying condition~(\ref{twolat}) can
either strengthen or weaken the band centre anomaly. They do not enhance
or suppress the modulation of the invariant distribution, but they can
widen or shrink the neighbourhood of the band centre where the invariant
distribution is significantly non-uniform.

\section{Conclusions}
\label{conclu}

In this paper we have studied the band-centre anomaly in the 1D Anderson
model with weak {\em correlated} disorder. Our perturbative analysis used
two essential tools: the Hamiltonian map approach and the continuum limit.
The Hamiltonian map approach interprets the spatial structure of
the electronic states in terms of the time evolution of a classical
parametric oscillator.
The dynamical evolution of the angle variable of this oscillator is
dictated by the random map~(\ref{fourth}). Replacing this map with a
corresponding stochastic differential equation is a crucial step that
allowed us to derive our analytical results and that required the
elaboration of the specific integration scheme~(\ref{scheme3}).
We obtained analytical expressions for the invariant distribution of the
phase variable and for the localisation length.
These results are valid for weak disorder with arbitrary correlations
and generalise the formulae obtained in~\cite{Tes12} for the case of
uncorrelated disorder.

When disorder is uncorrelated, the invariant distribution of the phase
variable, which is uniform for non-resonant values of the energy, becomes
modulated for energies lying in a neighbourhood of the band centre.
This modulation, in turn, generates a deviation of the inverse localisation
length from the values predicted by Thouless' formula.
In qualitative terms, this picture holds also when disorder displays spatial
correlations. From a quantitative point of view, however, the {\em size}
and {\em extension} of the resonance effect can be dramatically altered.
Two extreme cases are discussed in Sec.~\ref{exp_dec} and~\ref{oscexpdec}.
In the first case, the site energies exhibit positive correlations which
decay exponentially with the distance between sites. In this case the
resonance effect is suppressed upon increasing the correlation length $l_{c}$.
In the second case, the correlations between site energies also decrease
exponentially in magnitude, but oscillate between positive and negative
values. In this case, increasing the correlation length $l_{c}$ strongly
enhances the modulation of the invariant measure, which tends to a
sum of four delta peaks for $l_{c} \gg 1$.
Correspondingly, the difference between the inverse localisation length and
the value predicted by the formula derived by Izrailev and Krokhin
increases with $l_{c}$.
The specific long-range correlations analysed in Sec.~\ref{longrangedcor}
do not alter the modulation of the invariant distribution with respect to
the case of uncorrelated disorder, but can strengthen or weaken the
anomaly in a different way, i.e., they can enlarge or restrict the interval
of the energy in which the resonant effect is relevant.

In conclusion, we have shown that correlations of the disorder can alter
the band-centre anomaly very strongly and in a variety of ways. In particular,
specific correlations can suppress or magnify the resonance effect at
the centre of the energy band.

\section*{Acknowledgements}

The authors acknowledge support from the SEP-CONACYT (M\'{e}xico)
under grant No. CB-2011-01-166382.
I. F. H.-G. and F. M. I. also acknowledge VIEP-BUAP grant MEBJ-EXC12-G and
PIFCA BUAP-CA-169, while L.~T. acknowledges the support of CIC-UMSNH grant
for the years 2014-2015.

\end{document}